\documentclass[onecolumn,trackchanges]{aastex631}
\usepackage{graphicx}
\usepackage{amssymb}
\usepackage{epstopdf}

\newcommand{\xmm}{\textit{XMM-Newton}}

\shorttitle{Non-thermal $X$-rays from pulsation-driven shocks in Cepheids}
\shortauthors{F. Fraschetti et al.}

\begin{document}


\title{Non-thermal $X$-rays from pulsation-driven shocks in Cepheids}

\author{Federico Fraschetti}
\affiliation{Center for Astrophysics | Harvard \& Smithsonian, 60 Garden Street, Cambridge MA02138, USA;\\ Department of Planetary Sciences - Lunar and Planetary Lab, Tucson, AZ, 85721}

\author{Konstantina Anastasopoulou}
\author{Jeremy J. Drake}
\author{Nancy Remage Evans}
\affiliation{Center for Astrophysics | Harvard \& Smithsonian, 60 Garden Street, Cambridge MA02138, USA}

\begin{abstract}
Rapid X-ray phase-dependent flux enhancement in the archetype classical Cepheid star $\delta$~Cep was observed by XMM-Newton and Chandra. We jointly analyse thermal and non-thermal components of the time-resolved X-ray spectra prior to, during and after the enhancement. A comparison of the time scales of shock particle acceleration and energy losses is consistent with the scenario of a pulsation-driven shock wave traveling into the stellar corona and accelerating electrons to $\sim$ GeV energies and with Inverse Compton (IC) emission from the UV stellar background leading to the observed X-ray enhancement. The index of the non-thermal IC photon spectrum, assumed to be a simple power-law in the $[1-8]$ keV energy range,  radially integrated within the shell $[3 - 10]$ stellar radii, is consistent with an enhanced X-ray spectrum powered by shock-accelerated electrons. An unlikely $\sim$100-fold amplification { via turbulent dynamo} of the magnetic field at the shock propagating through density inhomogeneities in the stellar corona is required for the synchrotron emission to dominate over the IC; the lack of time-correlation between radio synchrotron and stellar pulsation contributes to make synchrotron as an unlikely emission mechanism for the flux enhancement. Although current observations cannot rule out a high-flux two-temperature thermal spectrum with a negligible non-thermal component, this event might confirm for the first time the association of Cepheids pulsation with shock-accelerated GeV electrons.
\end{abstract}


\section{Introduction}

Rapid time variability in X-ray flux has proven a valuable approach to constrain the physical properties of a variety of astrophysical sources, from jets in Active Galactic Nuclei \citep[e.g.,][]{Markowitz.etal:22}, black hole accretion discs \citep[e.g.,][]{Wilkinson.Uttley:09}, tidal disruption events \citep{Hampel.etal:22}, accreting neutron stars emitting Type-I X-ray bursts from the surface \citep{Degenaar.etal:18} or protostellar disk chemically affected by stellar flares \citep{Cleeves.etal:17}. Time variability has also enabled X-rays observations to probe the particle acceleration in shock-driven plasmas, as opposed to, e.g., magnetic reconnection: variability on a year time scale in the X-ray filaments of the supernova remnant Cassiopeia A was associated \citep{Uchiyama.Aharonian:08} with fast synchrotron cooling in strong magnetic field;  variability on a year time scale at inward shocks of Cas~A \citep{Sato.etal:18} allowed to constrain overdensities in shocked ejecta, small-scale dynamo processes and particle acceleration therein \citep{Fraschetti.etal:18}; a $\sim 3$-days drop in the plasma temperature behind the shock possibly generated by the outburst of the Nova V745~Sco was jointly observed by  Swift, Chandra, and NuStar \citep{Drake.etal:16b} and attributed to the leak of a substantial energy fraction in the form of cosmic-rays (protons and heavier ions); laboratory astrophysics experiments have identified via X-rays diagnostics the dynamo mechanism rapidly amplifying magnetic fields at shock waves \citep[e.g.,][]{Meinecke.etal:14}.  

The luminous pulsating Cepheid supergiants have shown in recent years a surprising X-ray variability correlated with pulsation phase. Combined XMM-Newton and Chandra observations have shown an unexpected 4-fold enhancement of the $(0.3 - 2.5)$ keV X-ray flux from the archetype Cepheid $\delta$~Cep \citep{Engle.etal:17} over $\sim 0.2$ in pulsation phase \citep[pulsation period $\sim 5.37$ days,][]{Engle.etal:17}; the enhancement is observed at the phase of the maximum radius. Chandra also detected a rapid increase of X-ray flux in $\beta$~Doradus at the phase of maximal radius \citep{Evans.etal:18}. However, lack of X-ray  enhancement at the same phase was reported from XMM-Newton for the Cepheid $\eta$~Aql \citep{Evans.etal:21}. 

A rapid enhancement in X-ray flux in Cepheids could be due to pulsation-driven compression of the magnetic field leading to reconnection, as in solar flares, or to the propagation of pulsation-driven shocks, either of which could plausibly heat gas to UV and X-ray emitting temperatures ($10^4$-$10^6$~K). The latter scenario was investigated in detail by \citet{Moschou.etal:20}, who showed that observed phase-dependent X-ray enhancements could be the result of pulsation-driven shock heating of ambient hot plasma. Here, we aim to build on that work by examining whether or not such shocks could also produce { efficiently a} population of accelerated particles and, potentially, non-thermal X-ray emission.

In considering possible particle acceleration in Cepheid atmospheric shocks, it is also useful to recall the circumstellar environments and radiatively-driven winds of early-type stars that are likely to be permeated by traveling shocks owing to the ``line deshadowing instability" \citep{Lucy.Solomon:70,Owocki:15}. \cite{Lucy.White:80,Lucy:82a,Lucy:82b} built a phenomenological model for the soft X-ray emission from gas heated by radiation-driven shock waves travelling into the gas at the base of the wind: the gas re-emits in the soft X-ray band via collisional excitation of ionized heavy ions. A key assumption is that the radiative cooling time is much shorter than the time between two subsequent shocks, that can therefore be considered isothermal. This mechanism based on radiation-driven shocks was revised to interpret the hard X-ray \citep{Chen.White:91a} and $\gamma$-ray emission \citep{Chen.White:91b} from OB supergiants as Inverse Compton (IC) emission from the UV field photons upscattered by shock-accelerated electrons.

In the \cite{Lucy.White:80} mechanism, the instabilities produce compressive waves that steepen into shocks propagating outward. In stars as cool as Cepheids, winds are not thought to be driven by resonance lines that might drive shocks via the line deshadowing instability; however, evidence of traveling shocks driven by pulsation within a number of Cepheids has been collected in recent years. \cite{Bohm-Vitense.Love:94} found that steep increases in emission-line fluxes $\sim 0.1$ phase before the peak of light for the Cepheid l~Carinae suggest emission from a chromospheric shock wave moving at speed greater than $100$~km/s, i.e., of the order of the escape velocity. \citet{Sasselov.Lester:94} applied coupled hydrodynamics and non-LTE radiative transfer modelling of chromospheric emission lines and concluded that Cepheids have chromospheres heated by acoustic or magnetic wave dissipation in addition to transient heating driven by the pulsation dynamics. \cite{Mathias.etal:06} found evidence of two consecutive shocks per pulsation period at X Sagittarii from analysis of  spectral lines obtained using the HARPS spectrograph. \cite{Nardetto.etal:06} found broadening of metal lines in RS Pup and suggested this was evidence of compressions or shock waves traveling through the stellar atmosphere.   

In this paper, we present a new analysis of the thermal and possible non-thermal components of the X-ray spectrum of $\delta$~Cep during the flux enhancements observed by Chandra and XMM-Newton  \citep{Engle.etal:17} in order to place constraints on non-thermal emission and the underlying energetic electron population. We also compare acceleration and energy loss time scales for electrons accelerated at the pulsation-driven shocks to identify plausible mechanisms of non-thermal emission. We find that the time scale of IC scattering of the UV photons from the stellar radiation field off the shock-accelerated electrons could potentially give rise to rapid non-thermal flux increases.

The structure of this paper is as follows: in Sect.\ref{sec:observations} the high- and low-state photon spectra are fitted with a thermal/non-thermal model and with a two-temperature model; in Sect.\ref{sec:timescales} the time scales for radiative cooling and electron shock-acceleration are compared with a variety of energy losses processes. In Sect.\ref{sec:discussion} the results are discussed and conclusion follows in Sect.\ref{sec:conclusion}.

\section{X-ray observations}\label{sec:observations}

\begin{figure}[ht]
\begin{center}
\includegraphics[scale=0.6]{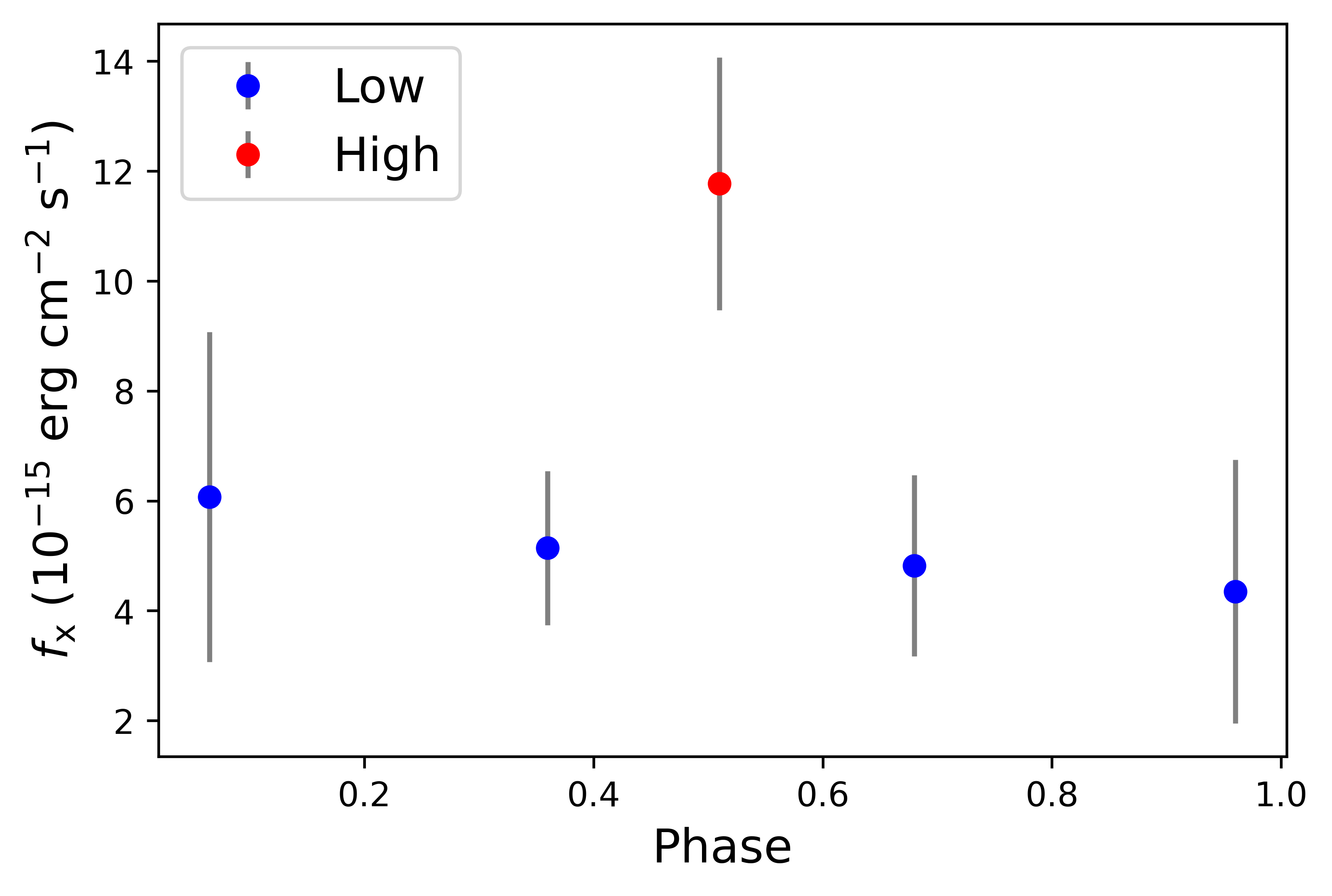}
\end{center}

\caption{The X-ray flux vs pulsation phase from the {\it XMM-Newton} observations of $\delta$~Cep used in this study, based on the fluxes and phases presented in \citet[][]{Engle.etal:17}. Phase=0 corresponds to maximum light.
We present the average flux and phase values of observations that were split in \citet[][]{Engle.etal:17}.
``Low" and ``High" $\rm{0.3-2.5\,keV}$ flux designations for the different observations are shown with red and blue dots, respectively, }. 
    	
\label{fig.phase_high_low}
\end{figure}

\cite{Engle.etal:14} analyzed the X-ray observations of $\delta$~Cep using a two-temperature thermal spectral model to calculate the flux. 
In order to obtain observational constraints on the possible presence non-thermal X-ray emission, we have performed a new analysis of the \emph{XMM-Newton} data, combining a thermal and a power-law spectral model.  We re-analyse the five  observations of \xmm{}, and fit separately the high-flux ($\rm{f_{0.3-2.5\,keV}>8\times10^{-15}\,erg\,s^{-1}\,cm^{-2}}$) and the low-flux ($\rm{f_{0.3-2.5\,keV}<8\times10^{-15}\,erg\,s^{-1}\,cm^{-2}}$) observations according to the flux measurements nomenclature of \citet[][]{Engle.etal:14,Engle.etal:17} but without splitting the single observations (see Figure  \ref{fig.phase_high_low}). Low and high flux observations are fitted separately to identify different physical mechanisms---thermal and non-thermal--- that could be imprinted in the spectra. Our analysis shows that, for a one-temperature thermal spectral model, the X-ray flux above $\sim 1$ keV is significantly in excess of the thermal flux and the spectrum of such a hard component can be fitted with a single power law in the range $[1-8]$ keV. We show in Table~\ref{tab.observations} the \xmm{} observation IDs and the exposure times corresponding to each of the three EPIC detectors after cleaning for background flares. { We have used the ephemeris reported in Table 1 of \cite{Engle.etal:14}, namely: $t(obs)= 2455479.905 + 5.366208(14) \times E$ days, where $t(obs)$ is the time of the observation, $2455479.905$ days is the time of a known past maximum light, $t_p=5.366208(14)$ days and $E$ is the epoch; the phase in Fig.\ref{fig.phase_high_low} is the decimal part of $E$.}

\begin{table}[ht]
	\centering
		\caption{Summary of $\delta$ Cepheid \xmm{} 
		{ listing observation ID, start and end date, and clean} exposure times in the MOS and pn cameras}
		\begin{tabular}{@{}lccrrr@{}}
			\hline
		Observation & Start date & End date & Exp pn & Exp MOS1 & Exp MOS2\\
		ID & & &(s) & (s) & (s) \\ 
		\hline
&&\underline{High-flux}&&&\\[5pt]
0603741001 & 2010-01-22 18:05:45& 2010-01-23 14:17:43
&69820& 72146 &72146  \\
\hline
&&\underline{Low-flux}&&&\\[5pt]
0552410401 &2008-06-05 14:26:02
&2008-06-05 21:53:07& 24685 &26112 &26107 \\
0603740901 &2010-01-20 18:04:11&2010-01-21 12:37:47& 49019 &56208 & 56483 \\
0723540301 &2013-06-28 06:11:55&2013-06-29 14:54:45& 103583 & 106436 & 106492 \\
0723540401 &2013-07-02 05:37:02&2013-07-03 08:00:22& 92020 &93613 & 93620 \\
\hline

			\end{tabular} 	
		\label{tab.observations}
		
		\smallskip
{}
			\end{table}

\subsection{Calibration and spectral extraction}

We reduced and analysed all EPIC archival observations for $\delta$~Cep using the \xmm{} Science Analysis System (SAS) v17.0.0. Calibrated event files were produced using the SAS tasks \textit{emchain}, and \textit{epchain} for the MOS and pn cameras respectively. Moreover, out-of-time event files were created for the pn camera using \textit{epchain} and\textit{withoutoftime=Y}, in order to subtract out-of-time events and correct for the charge transfer inefficiency.
The tool \textit{emtaglenoise} was used in order to flag noisy MOS CCDs at low energies \citep[][]{kuntz08}.
 
We filtered background flares by creating good time interval (GTI) files using the SAS task \textit{tabgtigen} with a cut-off of 2.5 counts/s for the MOS and 8.0 counts/s for the pn camera. 
We created a background light-curve for each detector, which was visually inspected to ensure the proper removal of the background flares. This resulted in filtered event lists for the pn and MOS cameras.

We then extracted spectra and calibration files for all the observations of $\delta$~Cep.
In order to extract the source and background spectra, we used the \textit{evselect} SAS task, and limited the patterns to singles and doubles for both EPIC pn and MOS event files. 
For the extraction of the background spectra, we selected a region nearby the source making sure we were not including any background sources.
We then extracted the Auxiliary Response Files (ARF) and the Redistribution Matrix Files (RMF) for the source spectra using the SAS task \textit{arfgen} and \textit{rmfgen}, respectively.
 
For the fitting of the spectra, we have used the XSPEC v12.11.1 software \citep{xspec}. In order to allow for chi-square fitting, we grouped the spectra in bins of 25 total counts.
For the high-flux spectra (see Table~\ref{tab.observations}) we fitted simultaneously the spectra from all EPIC detectors with all model parameters tied together. We also introduced a multiplicative constant, frozen to unity for pn and free for the MOS detectors, that was applied to each spectrum in order to account for residual calibration offsets between the different detectors.
For the low-flux spectra we used only the pn detector since the MOS detectors had very few counts. A multiplicative constant was introduced as well, but this time to account for differences in the long-term variability, or long-term residual variations in the calibration.
The line of sight Galactic absorption was in each case let free to vary.

In Table~\ref{tab.highflux} we report our spectral fitting results for the best-fit models of the high and low flux spectra respectively.
The multiplicative parameters show in all cases a difference of about 10-15\% between pn and MOS detectors. For the low-flux pn observations the multiplicative parameters do not differ in any case more than 4\%.

\subsection{High-flux spectrum}\label{sec:obs_high}

A two-component thermal model was used by \citet[][]{Engle.etal:14} to fit the X-ray spectrum of $\delta$ Cepheid.
Indeed, in the case of the high-flux observation, a best-fit single thermal component with temperature of $kT=0.18$\, keV does not provide a satisfactory fit to the spectrum ($\chi^2_\nu>1.6$) with residuals below 0.6\,keV and above 1.5\,keV. 
If a second empirical thermal component is introduced (model in XSPEC: \texttt{tbabs$\times$(apec+apec)}), the fit significantly improves, with $\chi^2_\nu=0.79$ and temperatures of $\rm{kT_1=0.12\,keV}$, and $\rm{kT_2=0.60\,keV}$
.

\begin{figure}[ht]
\begin{center}
\includegraphics[scale=0.32]{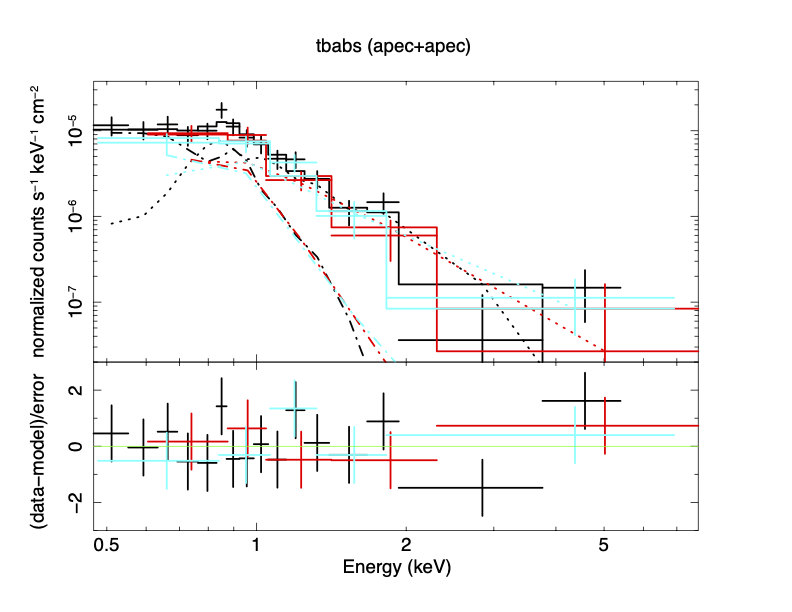}
\includegraphics[scale=0.32]{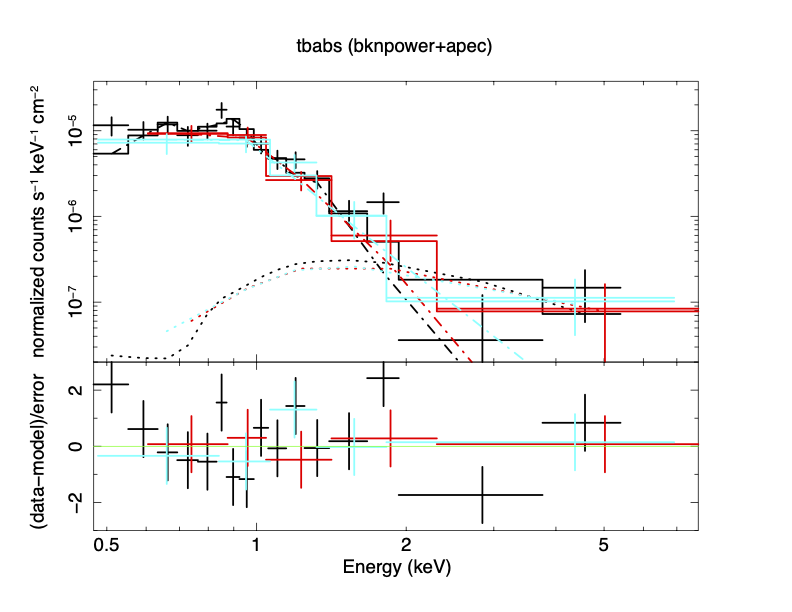}
\end{center}
\caption{Simultaneous fitting of all EPIC detectors (pn:black, MOS1:red, and MOS2:cyan) for the high-flux observation of $\delta$~Cepheid. Left: on the top panel we show the integrated X-ray spectrum, along with the best-fitting folded model consisting of two absorbed APEC components (dashed-dotted line and dotted line respectively), and on the bottom panel we show the fit residuals in terms of sigmas with error bars of size 1$\sigma$. Right: same as left with the best-fitting folded model consisting of an absorbed APEC component (dash-dotted line) and a power-law component (dotted line).
}     	
\label{fig.high}
\end{figure}

In order to explore the contribution of a possible non-thermal component at higher energies ($>$1.5\,keV) we introduced a ``broken power law'' model (model in XSPEC: \texttt{tbabs$\times$(bknpower+apec)}) where the power-law photon index 
$\Gamma$ above a certain energy break $E_{0}$ is positive (declining spectrum at large energy) and below $E_{0}$ it is frozen to a negative value in order to suppress the spectrum at soft energies
. The energy break and the power-law photon index $\Gamma$ for $E>E_{0}$ are frozen at $E_0 = \rm{0.8 \, keV}$ and $\Gamma = 2.0$ or $1.8$ or $1.6$, respectively. Such values follow from the time scale analysis in Sects.~\ref{sec:timescales} and \ref{sec:discussion} that supports a non-thermal emission originating from IC off the UV stellar radiation field. Small variations in the value of the $E_{0}$ ($\pm0.3$ keV) do not alter significantly the fitting results. The thermal plus power-law  model gives an acceptable fit of 
$\chi^2_\nu=1.27$ or $1.29$ or $1.32$ for $\Gamma = 2.0$ or $1.8$ or $1.6$ (see table \ref{tab.highflux}). These fits include all EPIC detectors (pn, MOS1, MOS2). The column density $N_H$ best-fit value of the two-temperature fit is not statistically different from the one of the thermal+power law fit
. 
From the best-fit model we find that the contribution of the non-thermal flux to the energy-integrated source flux in the 0.3-10\,keV band is as large as $\sim$30\% .
We present the results of the best-fitting models in the first four rows of Table~\ref{tab.highflux} (we also report the best-fit values for $\Gamma=1.8$ and $\Gamma=1.6$) and in Fig.~\ref{fig.high}.

\subsection{Low-flux spectrum}

\begin{figure}[ht]
\begin{center}
\includegraphics[scale=0.32]{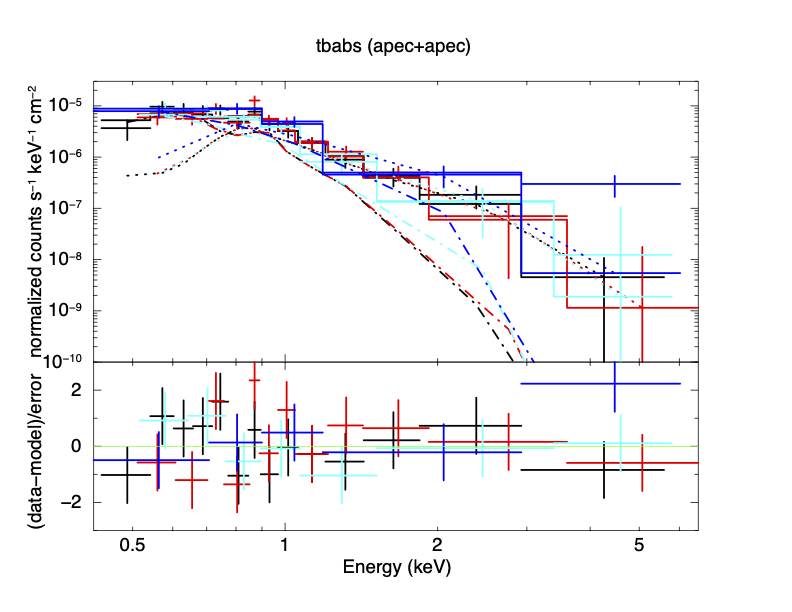}
\includegraphics[scale=0.32]{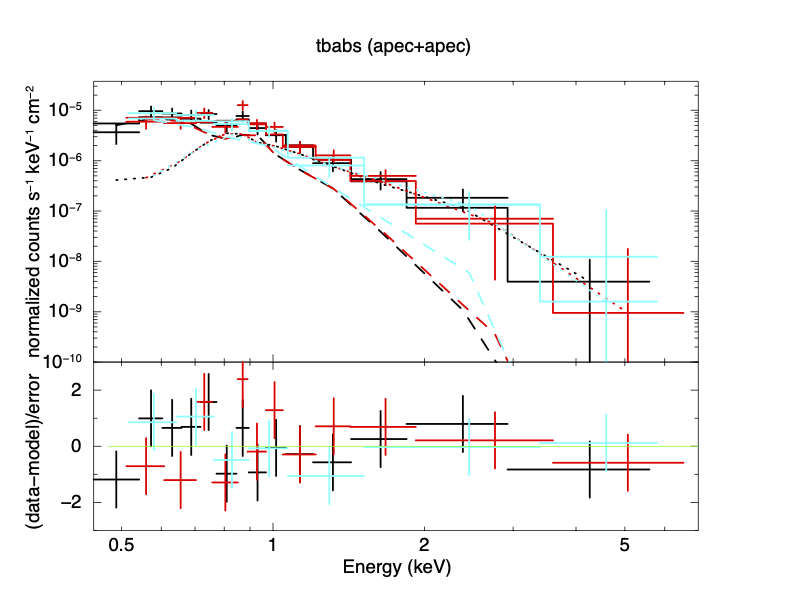}
\end{center}
\caption{Simultaneous fitting of all the low-flux pn observations shown in the colours black, red, cyan, and blue for OBSIDs 0723540301, 0723540401, 0603740901 and 0552410401, respectively. Left: on the top panel we show the time-integrated X-ray spectrum, along with the best-fitting folded model comprised of two thermal absorbed APEC components (dashed-dotted line and dotted line). The bottom panel shows the fit residuals in terms of sigma with error bars of size 1$\sigma$. Right: same as the left panel but 
excluding OBSID 0552410401 (blue datapoints).}     	
\label{fig.low}
\end{figure}

The same fitting method as in the high-flux spectrum case was followed for the low-flux spectrum. In this case we included only pn EPIC spectra of the four different low-flux observations (see Table \ref{tab.observations}) because in the MOS detectors the lower effective area leads to weaker signal that contributes less that 3 bins in each observation.
In the left panel of Fig.\ref{fig.low} we show the low-flux spectrum for all pn observations in different colours. We notice that observation 0552410401 (blue datapoints) appears to show a harder spectrum based on the single data point above 3\,keV. In fact, this particular observation exhibits higher flux values when split into different phases, as shown by the green points in Figure~3 of \citet{Engle.etal:17}: at phase $\sim 0.4$, during the rise phase, the flux is about 50\%\ higher than the low-flux observations (yellow datapoints of Figure~3 therein). Moreover, given the same normalization of the spectral peak, the pn flux at 5 keV in the high flux (black points in Fig.\ref{fig.low}) is $\sim 10^{-7}$ ct/s keV cm$^2$ and in the low flux (Fig.\ref{fig.low}, right panel, excl. 0552410401) the pn flux is about 10 times smaller.  
Therefore, the observation 0552410401 is likely not to belong with the low-flux observations, since it shows  harder spectrum and higher flux similar to that of the high-flux observation. For that reason we performed separate spectral fits including and not including observation 0552410401.

The best-fit model for both cases (including OBSID 0552410401: $\chi^2_\nu=1.12$, and excluding OBSID 0552410401: $\chi^2_\nu=1.07$) is given by a two-component thermal plasma model (last two rows of Table \ref{tab.highflux}).  Figure \ref{fig.low}, left  panel  shows the time-integrated spectra including OBSID 0552410401 (blue datapoints) and fitted with this two-component thermal model while the right panel shows the same fit excluding OBSID 0552410401. 
The non-thermal electrons are expected to be cooling down in the low-flux interval, as shown by the time scale analysis in Sect.\ref{sec:timescales}; thus, the power-law component is expected to be either reduced or suppressed in the low-flux spectrum relative to the high flux.
We have performed spectral fits with a thermal+power-law component following the same logic as for the high-flux observation.  
By including OBSID 0552410401, a fit with a given $\Gamma=2.0$ yields $\chi^2_\nu=1.30$ while excluding it yields $\chi^2_\nu=1.35$. 
From the best-fit models we find that the contribution of
the non-thermal flux to the energy-integrated source flux in the 0.3-10 keV band is similar to that high-flux case at $\sim$29\% in the case where we include OBSID 0552410401, and  $\sim$25\% when OBSID 0552410401 is excluded. 

Overall, the combined flux-spectrum analysis leads to conclude that a non-thermal component, although not completely suppressed in the transition between the high- and low-flux state, is strongly reduced. A non-thermal contribution might harden the low-flux spectrum via OBSID 0552410401. However, since it contributes only one datapoint ($\sim 5$ keV) with relative large error in energy, its statistical significance does not allow to firmly conclude that in the low-flux state the high-energy electron population has completely cooled.

\begin{table}[ht]
	\centering
		\caption{Best-fit parameters from spectral fitting of the high-flux and low-flux observations}
		\begin{tabular}{@{}lcccccccc@{}}
			\hline
		
		Model & $\rm{N_{H}}$&kT$_1$ & $\rm{norm_{kT_1}}$ & kT$_2$ & $\rm{norm_{kT_2}}$ & $\rm{norm_{\Gamma}}$ & $\chi^2$ (dof) & $\chi^2_{\nu}$ \\
		&$10^{22}$ atoms\,cm$^{-2}$ & keV & $\times10^{-4}$& keV & $\times10^{-5}$ & $\times10^{4}$  & &  \\
			\hline	
			&&&\underline{High-flux}&&&&&\\[5pt]
		
		 \texttt{tbabs$\times$(apec+apec)}&$0.79_{-0.24}^{+0.47}$ &$0.11\pm 0.04$ & $>35.0$ & $0.57_{-0.27}^{+0.74}$ &$2.4_{-1.2}^{+13.6}$&-& 15.0 (19) &0.79 \\	 [5pt]		 	\texttt{tbabs$\times$(bknpower+apec)} ($\Gamma$=2.0)& $0.85\pm 0.09$&$0.17\pm0.03$ &$8.7_{-5.4}^{+17.6}$&-& - &$1.2\pm0.6$  & 25.5 (20) & 1.27 \\[5pt]
		 \texttt{tbabs$\times$(bknpower+apec)} ($\Gamma$=1.8)& $0.86\pm 0.09$&$0.17\pm0.03$ &$8.9_{-5.4}^{+16.4}$&-& - &$0.9\pm0.5$  & 25.8 (20) & 1.29 \\[5pt]
		 \texttt{tbabs$\times$(bknpower+apec)} ($\Gamma$=1.6)& $0.87\pm 0.09$&$0.17\pm0.03$ &$8.9_{-5.5}^{+15.2}$&-& - &$0.6\pm0.4$  & 26.4 (20) & 1.32 \\[5pt]

		 \hline
		 &&&\underline{Low-flux}&&&&&\\[5pt]

	 \texttt{tbabs$\times$(apec+apec)}(incl. 0552410401)&$0.61_{-0.35}^{+0.53}$ &$0.15_{-0.06}^{+0.10}$ & $>2.3$ & $0.62_{-0.23}^{+0.85}$ &$0.6_{-0.2}^{+3.1}$&-& 33.58 (30) &1.12 \\	 [5pt]
	 \texttt{tbabs$\times$(apec+apec)}(excl. 0552410401)&$0.63_{-0.34}^{+0.67}$ &$0.14_{-0.06}^{+0.09}$ & $>2.4$ & $0.59_{-0.22}^{+0.75}$ &$0.7_{-0.4}^{+2.3}$&-& 28.0 (26) &1.07 \\	 [5pt]	
	 
	\hline

			\end{tabular} 	
		\label{tab.highflux}
		
		\smallskip
{The normalisation of the APEC component is in units of  $\frac{10^{-14}}{4\pi D^2}\int n_e n_H dV$ where $n_e$ and $n_H$ are the electron and hydrogen densities integrated over the volume V of the emitting region and D is the distance to the source in cm. The normalisation of the power-law is in units of $\mathrm{photons\,keV^{-1}\,cm^{-2}\,s^{-1}}$ at 1 keV. 
}
			\end{table}

\section{Time scales for thermal and non-thermal emission at pulsation-driven shocks} 
\label{sec:timescales}

In this section, we compare the time scale of electron acceleration at the pulsation-driven shocks with the time scale of the radiative cooling and the non-thermal energy loss processes relevant to relativistic electrons (Inverse Compton, synchrotron, bremsstrahlung, Coulomb losses) under the constraint that these processes 
deplete the high-energy electron population 
on a time scale shorter than the pulsation period $t_p$. The time scale of each process is related, in each separate subsection, to the properties of the wind or the propagating shock and compared with $t_p$.

\begin{figure}
	\includegraphics[width=6.4in]{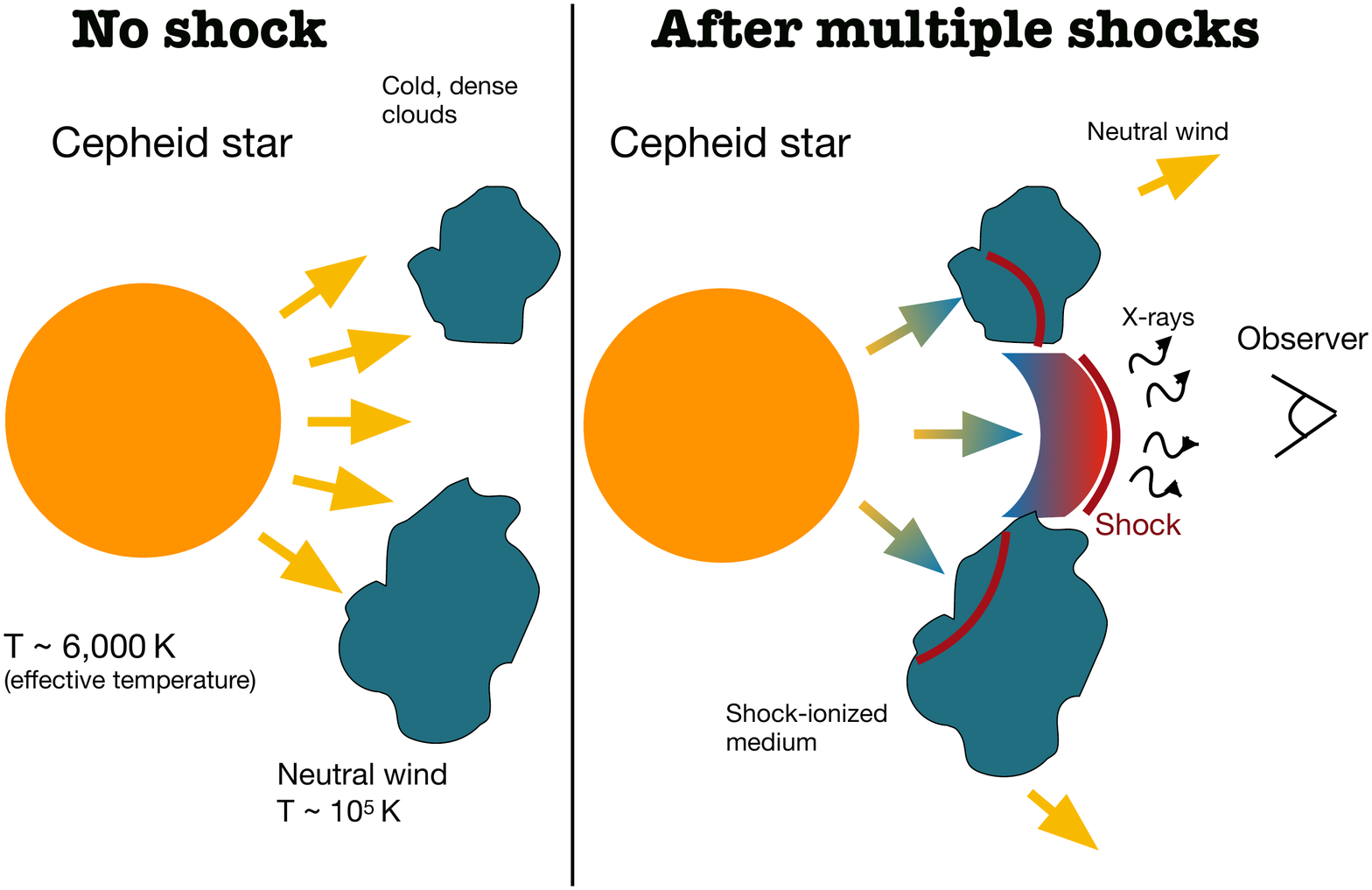}
    \caption{Cartoon illustration of the pulsation-driven shocks propagating into the pre-ionized circumstellar medium.
    }
 \label{fig:Cepheid}
 \end{figure}

The thermal spectra shown in Sect.\ref{sec:observations} reveal a shock-heated plasma at very high temperature far out from the star, much hotter than the typical effective temperature of a Cepheid ($T_{eff} \sim 6,000$ K). 
\citet{Moschou.etal:20} concluded that Cepheids exhibit two-component X-ray emission with shock waves being responsible for the phase-dependent variable emission (phases 0.2–0.6) and with a separate mechanism being the dominant source of emission for the ``quiescent" phases. In this scenario, the shock heating acts on this preexisting hot plasma 
only in locally more rarefied wind regions.
A sudden thermal X-ray emission, distinct from the continuum X-ray emission can be reconciled with a cold stellar wind only if a shock encounters a density inhomogeneity or undergoes strong time variability. { This possibility is also consistent with the low surface flux of $\delta$~Cep combined with fairly high temperature $\sim 1$ keV, as emphasized by \cite{Engle.etal:17}, that might suggest an emission localized to a small angular region.}
We envisage the scenario (see Fig.\ref{fig:Cepheid}) of a stellar pulse that in its expansion outward encounters inhomogeneities in the circumstellar medium that slows down the pulse, whereas the regions of the pulse expanding unimpeded by inhomogeneities might steepen into shock waves and heat/ionize the medium. As the stellar pulsation continues, multiple shocks cross the same circumstellar region that, after the passage of a ``first'' shock, encounter a mostly ionized coronal plasma, instead of the quiescent $\delta$~Cep neutral wind. A non-spherical distribution of pulse velocity might also originate deeper in the stellar atmosphere, { rather than being caused by the ambient medium. Cold ($\sim 40$ K) interstellar clouds detected in IR ($70$ and $160\, \mu$m) {\it Herschel} images around the Cepheids V Cen and RS Pup \citep{Hocde.etal:20} observationally support the scenario shown in Fig.\ref{fig:Cepheid}}.

{ Circumstellar envelopes (CSEs) around $\delta$~Cep (located at $2.4 \pm 0.1 \, R_\star$), and the Cepheid Polaris, were detected in the near infrared (IR) band K' ($1.9 - 2.3 \, \mu$m) via interferometry techniques using the Fiber Link Unit for Optical Recombination (FLUOR) instrument at the Center for High Angular Resolution Astronomy (CHARA) array \citep{Merand.etal:06}; CSEs were also detected around other Cepheids in the IR \citep{Kervella.etal:06} and mid-IR \citep{Gallenne.etal:21}, the latter for 13 stars ($29\%$ of their sample). The CSEs seem uncorrelated with the stellar mass loss, temperature, or with the presence of carbon monoxide \citep{Gallenne.etal:21}, and remain unexplained. Cold/hot dust composition of the CSE is ruled out by IR observations at $30\, \mu$m \citep{Hocde.etal:20}. A thin ($\sim 0.15\, R_\star$) shell of ionized gas was postulated \citep{Hocde.etal:20} as the origin of the IR emission (for a sample of $5$ Cepheids, not including $\delta$-Cep), in agreement with an expanding H$\alpha$ shell within the stellar atmosphere \citep{Gillet:14}. The presence of ionized material within a few $R_\star$ and H$\alpha$ profiles supports the scenario of shock waves driving the  ionization of the circumstellar region proposed here  (see Fig.\ref{fig:Cepheid}).}

{ In relation to the expected periodicity, the level of precision of observations for $\delta$-Cep and Polaris is insufficient for drawing conclusions on any asymmetry or time modulation of the CSE \citep{Merand.etal:06}; likewise, mid-IR excess,  available only at a specific phase, does not lead to any time-dependence conclusion. However, the change in the IR atmospheric opacity at $4.6 \, \mu$m due to the destruction-reformation cycle of carbon monoxide \citep{Scowcroft.etal:16} might be related to the stellar pulsation, and hence to the X-ray emission from electrons accelerated at shock waves that ionize the CSE in their outward propagation.}

\subsection{Radiative cooling}

The X-ray spectral fit shown in Sect.\ref{sec:observations} is consistent with a gas shock-heated to a temperature $T_{sh} \simeq 3 \times 10^6$~K from a shock wave travelling at a few hundred km/s: from the jump conditions, the  post-shock temperature scales as $T_{sh} = 3 m_p \mu V_{sh}^2/16 k \sim 1.5 \times 10^6 K (V_{sh}/(300 km\, s^{-1}))^2$, where $V_{sh}$ is the shock speed in the local upstream frame and the mean atomic mass in units of proton mass is benchmarked at $\mu = 1$.  
The X-ray light curve constrains the shock-heated gas radiative cooling time to be shorter than the pulsation period.

{The pristine and predominantly neutral wind gas of a Cepheid star is continuously ionized in local regions by travelling shocks, on a time scale very rapid compared with the wind expansion time scale.} Therefore, the ion density encountered by subsequent shocks passing through a post-shock region is much larger than the ion density in unshocked regions 
due to the freshly ionized neutrals and can be approximated with the pre-shock neutral density 
: $n_i^{sh} \sim n_H$. Thus, the source of the X-ray emission rarefies outward as $R^{-2}$. 
\cite{Moschou.etal:20} used 1D radial wind density profiles for various mass loss rates and determined the location $R$ of the X-ray source by intersecting these profiles with the value of the density needed for a radiative cooling time of 12 hours, to match the decay of the X-ray peak (Fig.7, lower panel, therein). Here we allow for a 3D distribution of cold partially neutral clumpy circumstellar medium that slows down shocks and a  highly ionized medium heated up by shocks that radiative cools on a time scale shorter than $t_p$ and that supplies the electrons accelerated by the shocks.
  
The cooling via collisional excitation of heavy ions by electrons 
has been calculated for the optically thin highly ionized plasma in the solar corona by, e.g., \cite{Landi.Landini:99,Colgan.etal:08} and can be approximated for temperatures in the range $10^6 < T < 10^7$ K 
by $ \Lambda(n_i, T) =  n_i n_H \, \tilde \Lambda(n_i, T_{sh}) =  n_i n_H \, 10^{-23}$ erg /cm$^3$s, where $\Lambda(n_i, T)$ is the cooling function, ${\tilde \Lambda(n_i, T_{sh})}$ is the radiative power per unit of volume (or radiative loss) and the global neutrality condition is assumed ($n_e \sim n_i$).  
 
In the case of OB supergiants \citep{Chen.White:91a} the pre-shock wind is mostly ionized with an average ion mass loss comparable with the neutrals mass loss in Cepheids ($10^{-6} \, M_\odot$/yr).

{The ion-to-neutral fraction in the unshocked wind of $\delta$~Cep is uncertain, but expected to be small due to the low effective temperature. However, the gas cooling is highest in the post-shock region, freshly ionized by the passage of the shock. Thus, the neutral gas behind the shock is assumed to be entirely ionized with a resulting $n_i$ much greater than in the pre-shock gas.} The neutral mass loss rate was estimated from $21$-cm line measurements of neutral hydrogen from circumstellar nebulae to be\footnote{For an optically thin wind of giants and supergiants in the spectral range G0-M5 \citep{Drake.Linsky:86} an upper limit to the steady time-averaged mass loss rate of the ionized component of the wind is $3.2 \times 10^{-10} M_\odot$/yr, assuming an outflow velocity of $35$ km/s (below the escape speed), $T=10^4$ K, and $R=41.6 R_\odot$. Lower surface temperature or wind speed lead to a  lower mass loss rate. This mass loss rate corresponds to a much smaller steady state ion density ($\sim 10^6$~cm$^{-3}$).} $\dot M \sim 10^{-6} \, M_\odot$/yr \citep{Matthews.etal:12}.  
Fourier decomposition of the wind speed in the numerical simulations by \cite{Moschou.etal:20} shows that at a radius $3\, R_\star$ speed fluctuations in the acceleration region settle further out ($R > 10\, R_\star$) to a coasting speed $\sim 100$~km~s$^{-1}$; only the post-shock region very close to the shock cools radiatively. Thus, the maximal speed $\sim 100$~km~s$^{-1}$ in the wind can be used to estimate the wind density in the shock region as follows. For the parameters listed above, the steady and optically thin mass loss corresponds to a neutral hydrogen number density
\begin{equation}
n_i (3\, R_\star) \simeq n_H (3\, R_\star) = \frac{\dot M}{4 \pi R^2 V (3\, R_\star) m \mu} = 
4.46 \times 10^9 cm^{-3} \frac{\dot M}{10^{-6} \, M_\odot/yr} \frac{100 \, {\rm km/s}}{V}  \left(\frac{8.4 \times 10^{12} \, {\rm cm}}{R} \right)^2   
\label{nH}
\end{equation}
where  
$R_\star = 40 R_\odot = 2.8 \times 10^{12}$ cm. The cooling time scale in the post-shock region is strongly reduced (with respect to the ion density in the unshocked wind) and at a distance $R$ from the star is 
\begin{eqnarray}
t_{rad} (R) &\simeq& \frac{3 \,n_i(R) \, k_B T_{sh}}{2\, \Lambda(n_i, T_{sh})} \nonumber\\
&\simeq& 0.72  \, {\rm d} \frac{T_{sh}}{3 \times 10^6 K} \frac{10^9 \rm{cm}^{-3}}{n_H(3\, R_\star)}\frac{10^{-23} \rm{erg \,cm}^3 /s} {\tilde \Lambda(n_i, T_{sh})} \left(\frac{R}{3 \, R_{\star}}\right)^2 \, ,
\label{eq:t_rad}
\end{eqnarray}
where the scaling $n_H \sim R^{-2}$ accounts for the coronal rarefaction and we have used $n_H \simeq n_i$ . Since the plasma continues its acceleration outward \citep{Moschou.etal:20}, the post-shock temperature is not appreciably decreased by the cooling. 

A comparison of $t_{rad}$ with $t_p \sim 5$~d (see Fig. \ref{fig1}) shows that, depending on several wind parameters, between $3\, R_\star$ and $10\, R_\star$ from the star the pulsation outruns the radiative cooling as $t_{rad}$ grows from $t_{rad} (3\, R_\star) \sim 0.72$ d to $t_{rad} (10\, R_\star) \sim 7 $ d $ > t_p$, due to the wind rarefaction, and the radiative cooling is suppressed at $10\, R_\star$. This rise of $t_{rad}$ provides a constraint on the location $R$ of the thermally emitting region in the $\delta$~Cep corona, although this condition remains uncertain due both to several poorly determined observational parameters, e.g., neutral and ion wind densities, and empirical models for the radiative cooling. Since the cooling of the post-shock region slows down as the shock moves outward, a larger fraction of the shock ram pressure can be transferred to electrons that can be heated/accelerated.

Note that we assumed that both the thermal and non-thermal $X$-ray emission \citep{Engle.etal:17} originate in the shock-heated plasma ($T_{sh} \sim 3 \times 10^6$ K). 
\cite{Moschou.etal:20} did not restrict the radiative cooling to the post-shock region and determined the radiative cooling time by using two likely average coronal plasma temperatures, both much colder than the post-shock plasma (for 18 out of 20 cases $T \sim 5 \times 10^5$ K, see Table 1 therein), that also adiabatically decrease ($T(r)$), 
thereby reducing $t_{rad} (R)$ by a factor $\sim 6$. 
At temperatures $10^5 - 10^6$ K
, the $\tilde \Lambda(n_i, T_{sh})$ is about 100 times larger than at $10^7$ K \citep{Landi.Landini:99,Colgan.etal:08,Moschou.etal:20}, thereby greatly suppressing the radiative cooling time scale\footnote{At such high temperatures ($\gg 10^4$ K), recombination of the gas is not efficient.},  
before it adiabatically cools outward into the range $10^4 - 10^5$ K where $\tilde \Lambda(n_i, T_{sh})$ is only $\sim$10 times larger than at $10^7$ K.

An opposite effect that increases $t_{rad}$ in the simulations by \cite{Moschou.etal:20} with respect to Eq.\ref{eq:t_rad} is the large wind speed.  
In the dense coronal plasma the shock layer, if estimated through the ion inertial length as heliospheric interplanetary shocks indicate, is remarkably thin: $c/\omega_{pi} (3\,R_{\star}) \sim 10^3 $ cm $ \sqrt{10^9 \, cm^{-3}/n_i^s (3\,R_{\star})}$, far shorter than the shock spacing, with a $t_p \sim 5$ days and speed $\sim 100$ km/s (these values lead to a shock spacing $\sim 4 \times 10^{12}$ cm); 
the shock transition layer is much thinner than any other scale involved in the system, including the compressed post-shock plasma region as found in the simulations 
by \cite{Moschou.etal:20}.
The maximal speed reached in the region behind each shock \citep[within thickness $\sim 10^{12}$ cm, see Figs. 3 and 5 in][]{Moschou.etal:20} is $\lesssim 200 $~km~s$^{-1}$, that leads to a smaller $n_H$ (see Eq.\ref{nH}) and a larger $t_{rad}$. However, this effect is at most of the order of a few whereas the decrease in $t_{rad}$ due to the temperature $T_{sh}$ 
can be as large as a factor 10, hence the very small $t_{rad}$ in \cite{Moschou.etal:20}.

\subsection{Electron acceleration at pulsation-driven shocks} \label{sec:acceleration}

We propose that the X-ray enhancement is powered by relativistic electrons accelerated at the shocks propagating out in the stellar corona. Generally, a higher acceleration rate at non-relativistic shocks is realized by a higher shock speed, a stronger magnetic field B or a magnetically oblique shock \citep{Jokipii:82,Drury:83}. Due to the rapid drop of the magnetic field with the radial distance from the star, i.e., $B \simeq R^{-3}$, the acceleration rate declines rapidly outward from the star, thereby constraining an X-ray emission on a scales shorter than $t_p$ to region very close to the star, i.e., within $10\, R_\star$. 

Electron acceleration at the propagating shock can be sustained diffusively up to large energies if the particle-particle collision mean free path is far greater than the hydrodynamical scale of the system. For fast electrons, the electron-ion collision frequency (much greater than the electron-electron collision frequency) scales as $\nu_{ei} \sim 10^{-9} \, n_i \, Z^2 \, \Lambda_{ei} E[eV]^{-3/2}$. For a $10$ keV electron, this  
leads to a collision mean free path within the shocked pre-ionized dense stellar corona of 
\begin{equation}
\lambda_{ei}= 3\, \frac{v}{\nu_{ei}} \simeq  10^{15} \, {\rm cm} \frac{10^9 \, {\rm cm}^{-3}}{n_i}     
\end{equation}
(where $v$ is the electron speed and we have used $\Lambda_{ei} \sim 20$) that largely exceeds all other length scales for the outward propagating shock, ensuring that throughout the  electron energy spectrum the collisionless regime holds. 
The population of shock-accelerated electrons ($> 10$ keV) is pitch-angle isotropic in the local upstream plasma rest frame due to the very high particle speed compared with the shock speed ($\sim 100$ km/s); thus, the scaling for the diffusive shock acceleration applies \citep{Drury:83}. Particle escape from the shock, competing with the acceleration and leading to a cut-off or no-power-law energy spectrum \citep[][]{Fraschetti:21},  or effects due to radial expansion \citep{Drury:11} are neglected herein; energy losses are detailed in Sect.\ref{sec:losses}. The acceleration time-scale, $t_{acc}$, for an electron of Lorentz factor $\gamma$ at a shock moving with speed $C_r$ in the stellar wind frame\footnote{For an expected stellar wind speed $\sim 30$ km/s within $10\, R_\star$, much smaller than the numerically determined shock speed ($\sim$ a few hundreds km/s at a distance of $3\, R_\star$, from \cite{Moschou.etal:20}), we can approximate $C_r$ equal to the speed in the stellar frame.} with an embedded magnetic field $B = |\mathbf{B}|$ can be approximated by \citep{Parizot.etal:06,Fraschetti.etal:18}
\begin{eqnarray} 
t_{acc} (\gamma) & \simeq & 1.83 \frac{3 r^2}{r-1}{D_0 (\gamma) \over C_r^2} \nonumber\\
&=& 2.4 \times 10^{-4} {\rm d}  \, \frac{ r^2 k_0}{r-1} \, \frac{0.01 G}{B(3\, R_\star)} \, \left(\frac{200 km/s}{C_r (R)}\right)^2 \frac{\gamma^2-1} {\gamma} \left(\frac{R}{3\,R_{\star}}\right)^3
\label{t_acc}
\end{eqnarray} 
where $r$ is the density compression at the shock, $D_0 (\gamma)$ is the spatial diffusion coefficient 
for an isotropic upstream turbulence, and $k_0 = D_0 / D_B$, assumed to be equal upstream and downstream \citep{Parizot.etal:06}, where $D_B$ is the Bohm diffusion coefficient at the electron Lorentz factor $\gamma$. 

As described by \cite{Chen.White:91a}, the background plasma in the post-shock region might loose most of its energy via radiative cooling until the dominant pressure terms become the magnetic or the non-thermal pressures. The non-thermal energy content at Mach number shocks $<10$ is typically $< 15\%$ of the total downstream energy (and of the upstream ram pressure) and it does not depend on the shock magnetic obliquity (i.e., angle between the upstream magnetic field and the normal to the shock surface), as an accurate analysis of in-situ measurements of interplanetary shocks found \citep{David.etal:22}. However, nova V745~Sco outbursts show a significant drop in plasma temperature within $\sim 10$ days, possibly due to a non-thermal energy content as large as $ 50 \%$ \citep{Drake.etal:16b}. 

The compression experienced by the energized electrons is larger at larger energies as the latter travel greater distances from the shock. The average displacement during $\Delta t \simeq 1$ day of a $1$ GeV 
electron diffusing around the shock can be as large as $\Delta x \sim \sqrt{2\, D_B \Delta t} \simeq 10^{11} $ cm 
, much greater than the presumed shock thickness $\sim c/\omega_{pi} (3\,R_{\star}) \sim 10^3 $ cm. {As a consequence, if the shock enters the radiative regime, the shock compression acquires a dependence on the particle energy and the estimate of $t_{acc}$ needs a correction beyond the scope of this paper.} 

As for the turbulence effect in the shocked region, the acceleration time scale varies considerably according to the geometry of the upstream magnetic field $\mathbf{B}$, i.e., if $\mathbf{B}$ is parallel or perpendicular to the shock normal (see Fig.\ref{fig1}). In the weak turbulence regime (far from Bohm regime), it holds that $k_0 \gg 1$ ($k_0 \ll 1$) for shocks in quasi-parallel (or quasi-perpendicular) geometry \citep[e.g.,][]{Fraschetti.Giacalone:12}; for strong turbulence, the total (average $+$ turbulent) magnetic field is nearly isotropic in configuration space leading to the Bohm regime ($k_0 \simeq 1$, blue curve in Fig.\ref{fig1}). For a quasi-perpendicular shock the acceleration rate has been long known to be higher \citep{Jokipii:82,Jokipii:87}, namely $t_{acc}$ shorter ($k_0 < 1$). Figure \ref{fig1} depicts the acceleration timescale as a function of electron kinetic energy for the case $k_0 =1$ (blue curve). The blue shaded (blank) region is accessible to quasi-perpendicular (quasi-parallel) shock geometry only: electrons can reach kinetic energy $> 1$ GeV at a quasi-perpendicular shock before the cooling dominates, according to loss processes described in Sect.\ref{sec:losses}. 

In Eq.~\ref{t_acc}, the ambient magnetic field strength is assumed to be dominated by the dipolar component, hence the scaling $B \simeq R^{-3}$ with the choice $B(3\, R_\star) = 0.01$G, that corresponds to the average surface magnetic field $0.43$ G inferred in a single spectropolarimetric observation with ESPaDOnS by \cite{Barron.etal:22}. This magnetic field  squeezes the region of particle acceleration close to the star before losses outpace acceleration, in contrast with the $B \simeq R^{-2}$ scaling in \cite{Chen.White:91a} where the acceleration time increases at a slower rate with the radial distance. 
A dominant monopole scaling $B \sim R^{-2}$ instead of the dipole used here would not change the conclusion in Fig.\ref{fig1}.

Hydrodynamic simulations \citep{Moschou.etal:20} show that the Mach number of the travelling compressions in the local plasma frame, approximately coinciding with  the stellar frame as the quiescent wind speed is only $\sim 30$ km/s, spans the range 1 to 10 very close to the star surface ($\sim 2\, R_\star$) and keeps increasing out to $\sim 40\, R_\star$. In a circumstellar medium with magnetic field $B (3 \,R_\star) \sim 0.01$ G and ion density $n_i = 10^9$ cm$^{-3}$ \citep[the models of][spanned values of ion number density between $10^6$ and $10^9$ cm$^{-3}$]{Moschou.etal:20}, the Alfv\'en speed is $\sim 1$ km/s, much smaller than the local sound speed \citep[$\sim 500$ km/s at the shock-heated regions][]{Moschou.etal:20}. Thus, the expected spectral slope of energized electron is approximately determined by the sonic Mach number radial profile as calculated in \cite{Moschou.etal:20} and no correction due to the magnetic field introduced herein is needed.

Morevoer, the hydrodynamic simulations in \cite{Moschou.etal:20} suggest an increasing shock speed $C_r (R)$ during the first phase of the shock propagation. Guided by these simulations, in the two snapshots of Fig.\ref{fig1} calculated at $R=3$ and $10\, R_\star$, we assigned to $C_r$ the two values 200 km/s to 400 km/s and compared $t_{acc}$ with the losses time scales described in the following section at these two locations. Further out, the decrease of the $B-$field with radius reduces the efficiency of the shock acceleration so that, at a given electron energy, losses prevail outward. In addition, the shock speed in the local plasma wind frame depends on the assumed stellar neutral mass loss, that is highly uncertain.

\subsection{Energy losses in the leptonic scenario}\label{sec:losses}

In this section we consider various processes of energy loss for the relativistic electrons accelerated as described in Sect. \ref{sec:acceleration} to a Lorentz factor $\gamma$ by a single pulsation-driven shock and compare the rates of energy loss with the acceleration time scale $t_{acc}$, the thermal radiative cooling time scale $t_{rad}$ and the stellar pulsation period $t_p$. Non-thermal emission is an unambiguous signature of the presence of relativistic electrons that can be transported far into the outer corona only by traveling shocks. 
The energy loss time scales increase rapidly with the distance $R$ and exceed $t_p$ only within a few stellar radii from the Cepheid star; however, losses can balance $t_{acc}$ only after the time necessary for the shock to energize the electrons up to the $\sim$GeV range. Time-dependent effects on the electron energy spectrum due to the continuous process of acceleration and energy loss are neglected here.

\subsubsection{Inverse Compton Scattering} 

We propose that the non-thermal enhanced X-ray emission in $\delta-$Cep  originates from UV ($\sim 1 -100$ eV) stellar wind photons upscattered into X-ray band by the IC over the shock-accelerated electrons, as proposed by \cite{Chen.White:91a} for OB supergiants. The target UV photon population within the photosphere/corona is assumed to be isotropic in the frame of the plasma flowing downstream of the shock, i.e., Doppler effect is negligible for a shock speed $C_r \simeq 200 \,{\rm km/s}$. The population of shock-accelerated electrons is also isotropic in the plasma frame at that shock speed as pointed out in Sect.\ref{sec:acceleration}.

In the Thomson limit ($4\varepsilon_0 \gamma/ m_e c^2 \ll 1$, where $\varepsilon_0 = 5$ eV is assumed as the UV initial photon energy), the power $P^T_{IC}$ emitted by a single electron is 
\begin{equation}
P^T_{IC} =  
 {4\over 3} \sigma_T c \gamma^2 \, U_{rad},
\label{P_IC_Thom}
\end{equation}
where $U_{rad}$ is the photon energy density integrated over all frequencies, 
 $\sigma_T$ is the Thomson cross-section and $c$ the speed of light in vacuum. 
The IC loss time scale in the Thomson limit at distance $R$ from the stellar surface, using Eq. \ref{P_IC_Thom}, is 
\begin{eqnarray}
t^T_{IC} (\gamma) &=& \frac{m_e c^2 \gamma}{P^T_{IC} } \, \left(\frac{R}{R_{\star}}\right)^2 \nonumber\\
&=& \frac{ 1.3 \times 10^3 \, {\rm d}}{\gamma } \left( \frac{6,000\, K}{{ T_{\star}}} \right)^4 \left(\frac{R}{3 R_{\star}}\right)^2 .
\label{t_loss_Thom}
\end{eqnarray}
where we have used for the $\delta$~Cep stellar surface temperature $ T_{\star} \sim 6,000$ K and $U_{rad} = \sigma T_{\star}^4 /c$ where $\sigma$ is the Stefan-Boltzmann constant, and the factor $({R}/R_{\star})^2$ accounts for the scaling of $U_{rad}$. The decrease of $t^T_{IC} (\gamma)$ with the electron kinetic energy is shown by the orange curve in both panels of Fig.\ref{fig1}. This curve intersects $t_{acc}$ and $t_p$ (black line) at the smallest energy (GeV) making it the most efficient loss process.

It is noteworthy that although in this scenario the X-ray non-thermal IC emission is assumed to originate from the same source as the thermal emission (shock region), the IC can be emitted by the energetic electrons escaped upstream from the shock, as well as those advected downstream of the shock, and interacting with the cool UV stellar wind photons ($6,000$ K). The scenario by \cite{Chen.White:91a} for OB supergiants also assumed that the sources of non-thermal/thermal emission are spatially coincident. However, the stellar wind in \cite{Chen.White:91a} is assumed to be isothermal in the circumstellar region travelled by the shocks
, in contrast with the radial scaling of the photon energy density $U_{rad}$, i.e., of the temperature, assumed here.

In the extreme Klein-Nishima limit ($4\varepsilon_0 \gamma/ m_e c^2 \gg 1$), for the monochromatic 
target photon field, 
the power $P^{KN}_{IC}$ emitted by a single electron is given by \citep{SChlickeiser:09,Fraschetti.Pohl:17b}
\begin{equation}
P^{KN}_{IC}    
 \simeq  {3\over 8} \sigma_T c (m c^2)^2 {n_0\over \varepsilon_0} \left[{\rm ln} {4 \varepsilon_0 \gamma\over m_ec^2} - {11\over 6}\right] ,
\label{P_IC_NK}
\end{equation}
where we have used $n_0 =  \sigma T_\star^4 /c \, \varepsilon_0$. The IC loss time scale, using Eq. \ref{P_IC_NK}, is 
\begin{eqnarray}
t^{KN}_{IC} (\gamma) &=&  \frac{m_e c^2 \gamma}{P^{KN}_{IC} } \, \left(\frac{R}{R_{\star}}\right)^2 = \frac{\gamma}{ {3\over 8} \sigma_T  m c^3 {n_0\over \varepsilon_0} \left[{\rm ln} {4 \varepsilon_0 \gamma\over m_ec^2} - {11\over 6}\right] } \left(\frac{R}{R_{\star}}\right)^2  \nonumber\\
&=&     4.5 \times 10^{-7} d\, \frac{\gamma  }{{ \left[{\rm ln} [3.9 \times 10^{-5} \, \gamma] - {11\over 6}\right] }  }  \left( \frac{6,000\, K}{{ T_{\star}}} \right)^4 \left(\frac{R}{3 R_{\star}}\right)^2 \,. \label{t_loss_NK}
\end{eqnarray}
Figure \ref{fig1} shows that the IC losses in the KN regime are ruled out, due to an electron energy cut off at energies too low at $R = 3\,R_\star$ for the KN regime.

\subsubsection{Synchrotron losses} 

The relativistic electrons also undergo synchrotron loss at the shock, that might contribute to the non-thermal emission. Assuming spatial 3D isotropy of the upstream magnetic field $B$, the magnetic strength is compressed  downstream to $0.83\, r \,B$, where the factor $0.83$ is due to the upstream isotropy \citep{Parizot.etal:06}. The synchrotron loss time scale can be written as 
\begin{equation}
t_{syn} (\gamma)=  \frac{10^8 d}{ \gamma} \left(\frac{0.01 G}{0.83\, r \,B(3\,R_\star)}\right)^2 \, \left(\frac{R}{3 \,R_{\star}}\right)^6 \, . \label{t_loss_sync}
\end{equation}
Figure \ref{fig1} shows that for $B(3\,R_\star) = 0.01$ G the synchrotron time scale becomes comparable with $t_p$ only in the $E\sim$ TeV energy range and close-in  ($\sim 3 R_\star$). If the magnetic field is locally much stronger both upstream and downstream (due to, e.g., plasma instabilities), for example $B(3\,R_\star) = 1$ G (Fig.\ref{fig2}), the Alfv\'en speed becomes comparable with sound speed thereby reducing the Mach number 
and synchrotron cooling dominates over the IC cooling off the UV stellar field; however, in such a large upstream B-field, the acceleration rate is also enhanced and the electron energy cut off, due to synchrotron cooling, increases to $\sim 5 $ GeV, higher than in the IC-dominated case. 

Another possible scenario of enhanced B is 
based on X-ray emission from the clumps, heated by the shocks, that are likely to cover a broad range of spatial scales within the inner corona. 
Such inhomogeneities can trigger an amplification of the magnetic field downstream of the shock due to vorticity generation as first demonstrated in MHD simulations by   \cite{Giacalone.Jokipii:07,Inoue.etal:12} and analytically described in terms of structure of upstream clumpiness by \cite{Fraschetti:13}. In the latter model the B-saturation strength depends on the radius $R_c$ of the  clumps and on the thickness $\ell_F$ of the outer layer where the gas density gradient is non-vanishing \citep{Fraschetti:13}; if $\ell_F/C_r \ll 1$ day, the amplified downstream magnetic field can saturate at a turbulent strength $\delta B$ far higher than the seed field $B(3\, R_\star)$, consistently with the $\delta$-Cep observations. Such an amplification process was suggested \citep{Fraschetti.etal:18} to explain time variability of X-ray flux at Cas-A supernova remnant \citep{Sato.etal:18}. The saturation turbulent field scales with the shock Alfv\'en Mach number $M_A$ as $\delta B / B(3\,R_\star) \simeq M_A \sqrt{2\,R_c/\ell_F}$ \citep{Fraschetti:13}. For $M_A \sim 30$ and $R_c/\ell_F \sim 50$, the magnetic field is $\sim$300-fold enhanced and the synchrotron cooling overruns the IC cooling off the UV stellar field (see Fig. \ref{fig2}): 
\begin{equation}
t'_{syn} (\gamma)=  \frac{7.3 \times 10^6 d}{ \gamma} \left(\frac{1 G}{\delta B}\right)^2 \, \left(\frac{R}{3 R_{\star}}\right)^6 \, .
\label{t_loss_sync_amp}
\end{equation}

It is noteworthy that, if the B-field amplification is caused by medium inhomogeneity, the synchrotron cooling time is shortened whereas the acceleration time scale, that depends on the upstream field, is unchanged. 

\subsubsection{Bremsstrahlung} 

In the dense coronal environment, even in the shock-compressed region, the density of ions generated by the ionization from the passage of a previous shock is at most of order $n_i (3 R_\star) \sim 10^9$ cm$^{-3}$ (see discussion above). In such a high density environment, the bremsstrahlung energy loss time scale due to collision of the energetic electrons of kinetic energy $E$ with thermal protons in a fully ionized hydrogen plasma or with thermal electrons is given by \citep[Eq.s 7 to 9][]{Haug:04} 
\begin{eqnarray}
-\left(\frac{dE}{dt}\right)_{brem} =  m_e c^3 \, n_i \left(\alpha \, \frac{3}{8\pi} \sigma_T \,  \frac{8 \gamma^2}{ \gamma^2 - p^2}A_p(E) + A_e(E) \right) ,\nonumber\\ 
t_{brem} (E)=  \left(\frac{1}{E} \frac{dE}{dt_{brem}}\right)^{-1} \left(\frac{R}{R_{\star}}\right)^2
\label{tbrem}
\end{eqnarray}
where $p = \sqrt{\gamma^2 -1}$ and the factors $A_p(E)$ and $A_e(E)$ describing the target particle, proton and electron respectively, are defined in the Appendix. 

In Eq.\ref{tbrem} we have neglected admixtures of heavier ions and considered only electron-proton bremsstrahlung.

Figure \ref{fig1} shows that for the ion number density expected in the stellar corona $t_{brem}  \gg t_p $ within the kinetic energy range $(1- 10^9)$ keV both at distance $3$ and $10 R_\star$ from the star; thus, denser plasma regions in the inner corona cannot power bremsstrahlung emission at sufficiently high rate to explain the X-ray enhancement \citep{Engle.etal:17}. In the tenuous outer stellar wind regions \citep[with shock-generated ion density $n_i \sim R^{-2}$,][]{Moschou.etal:20}, the bremsstrahlung loss rate for the accelerated electrons decays as the shock travels outward in the stellar corona.  
Only in an extremely dense wind ($n_H \sim 10^{10} $ cm$^{-3}$) can bremsstrahlung emission exceed the IC. However, such large plasma density also suppresses the radiative cooling time scale to only a few tens of minutes and the particle acceleration in a much cooler, denser and unstable downstream condition might not proceed adiabatically.  

\subsubsection{Coulomb losses} 

\begin{figure}
	\includegraphics[width=6.4in]{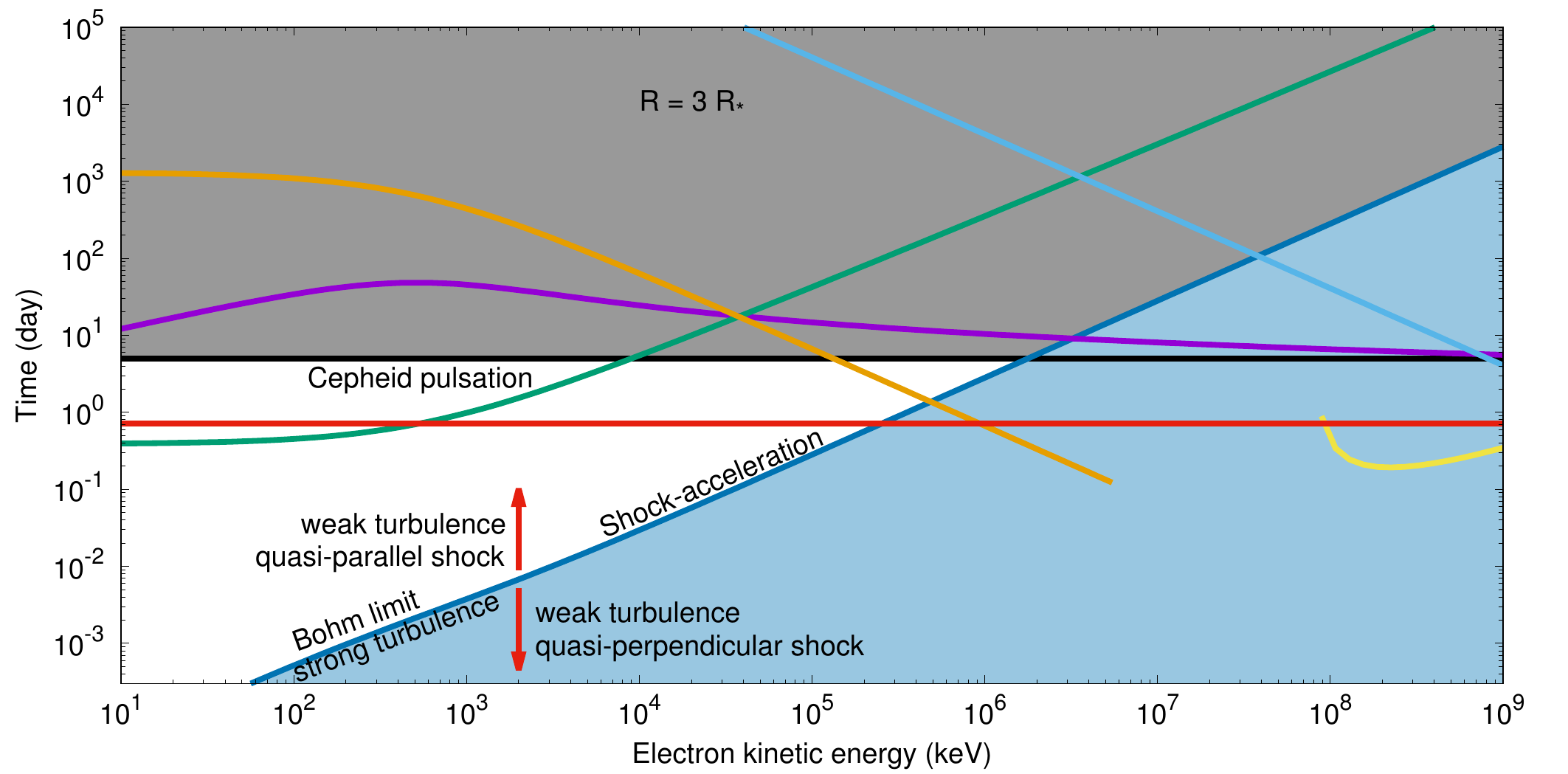}
	\includegraphics[width=6.4in]{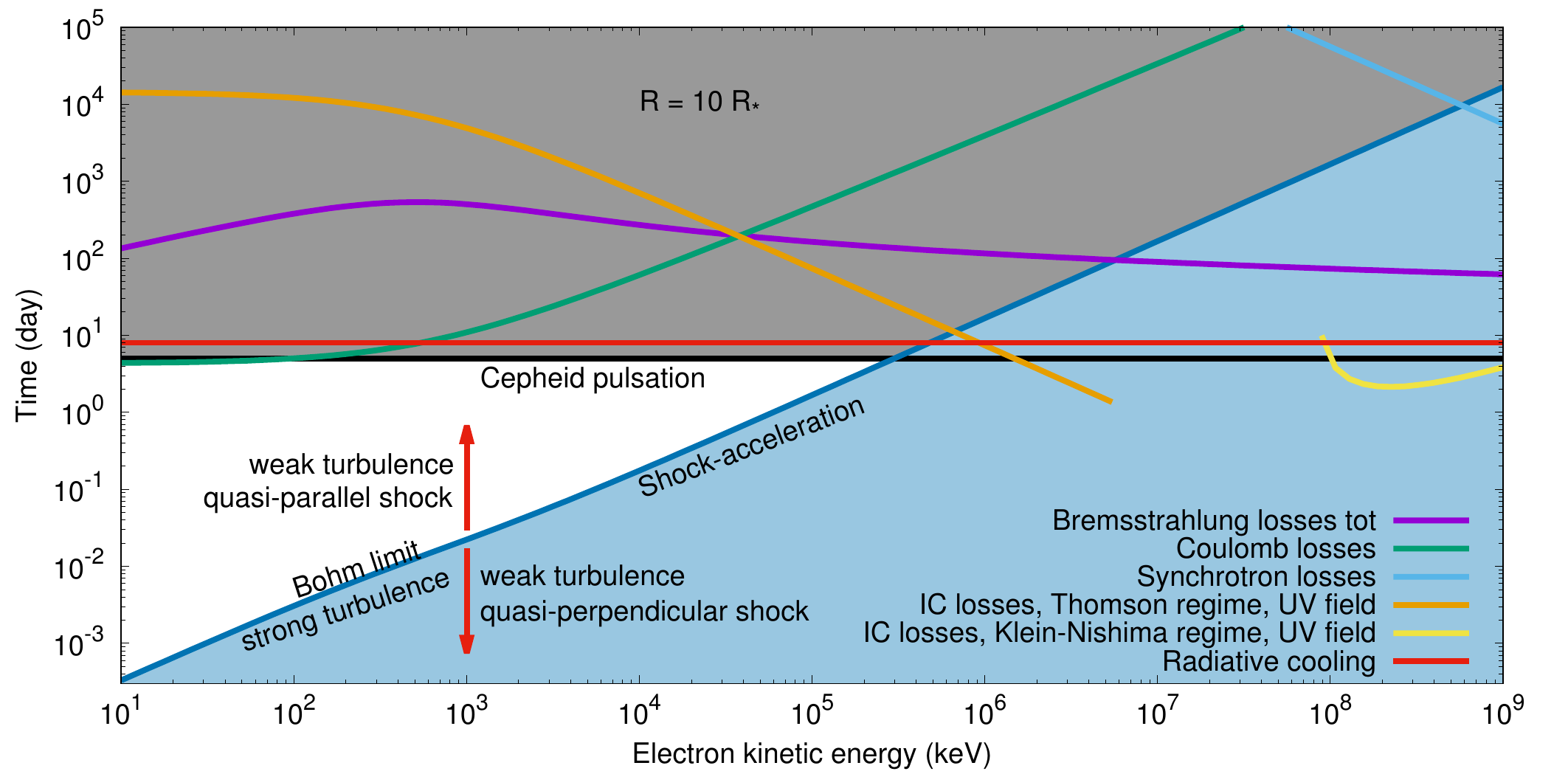}\\
    \caption{For energetic electrons accelerated at a shock (density compression $r=4$) travelling through the stellar corona, the time scales for the non-thermal energy losses (bremsstrahlung $t_{brem}$, IC $t^T_{IC}$ and $t^{KN}_{IC}$, Coulomb $t_{Cou}$, synchrotron $t_{syn}$), non-radiative shock acceleration $t_{acc}$ (in blue Bohm regime, or strong magnetic turbulence, i.e., $k_0 = 1$) and thermal radiative cooling $t_{rad}$ are compared with the estimated $\delta$~Cep pulsation period. The Bremsstrahlung time scale includes both e-e and e-p collisions \citep[][Eq. 9 therein]{Haug:04}. The shaded blue (blank) area below (above) the blue curve corresponds to the case $k_0 < 1$ ($k_0 > 1$), i.e., weak turbulence at quasi-perpendicular (quasi-parallel) magnetic obliquity. The grey shaded area above the black line corresponds to times longer than $t_p$ (black horizontal line), thus ruled out by X-ray observations. Top: $R=3\,R_{\star}$ with shock speed $C_r = 200$ km/s \citep{Moschou.etal:20}, $n_i(3\, R_\star) = 10^9$ cm$^{-3}$, $n_H(3\, R_\star) = 10^9$ cm$^{-3}$ and $B(3\, R_\star) = 0.01$ G. Bottom: Same as Left top with $R=10\,R_{\star}$ with shock speed $C_r = 500$ km/s and scaling of $n$, $n_H$ and $B$ as described in Sect.\ref{sec:losses}. 
    }
 \label{fig1}
 \end{figure}

\begin{figure}
	\includegraphics[width=6.4in]{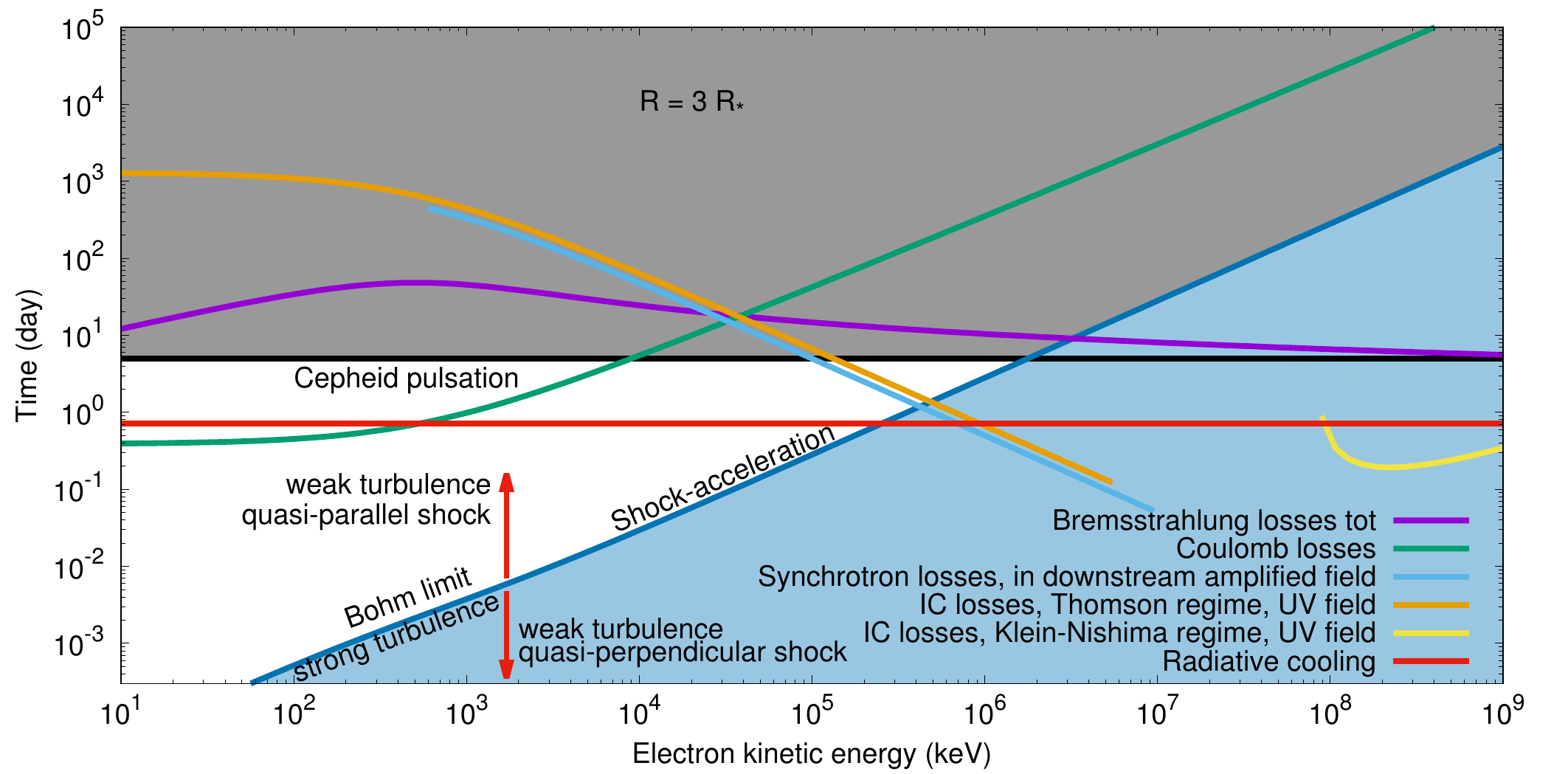}
    \caption{ 
    Same as Fig.\ref{fig1} with an amplified magnetic field $\delta B = 300 \, B(3\, R_\star) = 3$ G.}
 \label{fig2}
 \end{figure}

Although a shock ionizes most of the stellar wind gas that it encounters, subsequent shocks are likely to encounter a comparable neutral component. Thus, the rate of energy loss in the ionization of neutral hydrogen should be considered as well. The time scale for Coulomb losses can be expressed as \citep[][Eq. 19.1]{Longair:94}   
\begin{equation}
t_{Cou} (\gamma) \simeq 7.5 \times 10^2\, {\rm d} \frac{\gamma}{ 3\, {\rm ln}  + 19.8} \frac{10^9 \, cm^{-3}}{n_H (3\,R_\star)} \left(\frac{R}{3\,R_{\star}}\right)^2.
\label{tion}
\end{equation}
The Coulomb time scale is shown in Fig.s \ref{fig1} and \ref{fig2} in green to exceed all others listed above in the energy range of interest.

\section{Discussion} \label{sec:discussion}

\subsection{X-rays high- and low- flux state relation with the UV flux}

The observations of the X-ray enhancement in $\delta-$Cep are consistent with a physical model that combines a single thermal with a non-thermal component that cools on a time scale comparable with the stellar pulsation period. The observations analysed herein do not rule out that a non-thermal component might still be  present during the low-flux interval and did not completely dissipate during the high-flux, despite being  largely reduced. { The rate of ejection of the} shocks heating the plasma in the rarefied regions might be too high for the  non-thermal population to cool completely, as illustrated in Fig.\ref{fig1}. On the other hand, { a two-temperature model, similar to the one used in \cite{Engle.etal:17}, provides a good fit with no need of a} non-thermal component; however, the origin of the two distinct temperatures remains to be determined. 

{ The X-ray enhancement is observed during the phase of maximum angular radius whereas, based on photospheric observations, this particular phase of pulsation is rather quiet. Evidence of shocks from the far-UV peak is collected \citep{Engle.etal:17} on the ascending branch of the light curve, immediately after the radius minimum, e.g., $\sim 2.7$ days before the X-ray enhancement. Such a delay of a fraction of $t_p$ suggests that the radiative cooling occurs earlier within the pulsation period than the IC-cooling of the accelerated electrons; we find a comparable delay in Fig. \ref{fig1} (top panel) at $R=3\,R_\star$ between $t_{rad}$ (red curve) and the IC-cooling condition $t_{acc} = t_{IC}^T$, denoted by the intersection of the blue and orange curves. Our time-scale estimate shows that this process is constrained to occur within $10\, R_\star$ (see Fig. \ref{fig1}, bottom panel), and that the distance of $3\,R_\star$, close to the ionized shell postulated by \cite{Hocde.etal:20}, is consistent with the observed delay between X-ray flash and FUV peak.}

The scenario proposed here of an IC emissivity off the stellar UV background to generate the non-thermal component and radiative cooling from the shock-heated gas to produce thermal emission is consistent with the expected parameter for star and circumstellar medium. We discuss below some implications of a non-thermal scenario and a support from radio counterpart from VLA.

\subsection{Energetic electron cutoff}

A comparison of $t_{acc}$ with $t^T_{IC}$ and $t_{rad}$ in Fig. \ref{fig1} (blue, orange and red, respectively), shows that, due to the short radiative cooling time ($\sim$ a few days), the post-shock plasma approaches a radiative regime roughly at the same energy scale of the crossing of $t_{acc}$ with $t^T_{IC}$, that marks the inefficient acceleration compared with losses. 
The cutoff in electron energy is therefore provided by the condition \begin{equation}
t_{acc} (\gamma) = t^T_{IC} (\gamma) \quad \rightarrow \quad {\bar \gamma}^2 \simeq 2.0 \times 10^5 \, \left( \frac{6,000\, K}{{ T_{\star}}} \right)^4  \, \left(\frac{C_r (R)}{200 km/s}\right)^2 \, \frac{B(3\, R_\star)}{0.01 G} \, \frac{r-1}{r^2 k_0}  \frac{3 R_{\star}}{R} 
\label{eq:balance}
\end{equation}
or at $\bar E \simeq 0.2$ GeV at $R = 3 R_\star$ and decreasing with $R$. Closer to the star, the radiative cooling time scale within the shock-heated plasma is shorter than the pulsation period. At larger distance from the star, the wind rarefaction leads to an increase of the radiative cooling time scale that becomes longer than the pulsation period. Such a transition can be used as a diagnostic to determine the distance from the star of X-ray enhancement region, the local plasma density, hence to constrain the stellar mass loss. 

\subsection{Non-thermal component: energy range and spectral index}

The spectral fit of the X-ray flux rapid enhancement in the range [0.4-8] keV performed herein shows a significant excess over the thermal emission and reveals the presence of a non-thermal power component along with the soft (thermal) X-ray. 

The lower limit energy $0.8$ keV used in the non-thermal spectral fit in Sect.\ref{sec:observations} is based on the assumption that the non-thermal power-law is produced by IC off the UV stellar photon radiation field (with a benchmark photon energy $\varepsilon_0 = 5$ eV). On the other end, the balance acceleration/energy losses for the electrons energized at the pulsation-driven shock sets a maximal electron energy as a function of the distance from the star ($\gamma \gtrsim 100$ from Eq. \ref{eq:balance}). The two conditions above read for the observed photon energy $E$ \citep{Rybicki.Lightman:86}
\begin{equation}
    4\, \gamma_1^2 \varepsilon_0  \ll E  \ll 4\,\gamma_2^2 \, \varepsilon_0
    \label{eq:gamma_lim}
\end{equation}
where the single electron Lorentz factors $\gamma_i$, with $i=1,2$, are the minimum and maximum values discussed above. Equation \ref{eq:gamma_lim} leads to the constraints $80$ eV $\ll E \ll 200$ keV, hence the assumed lower limit $E=0.8$ keV for the spectral fit. It is noteworthy that the electron spectrum is assumed not to change both $\gamma_2$ and spectral index while the shock swipes through the stellar wind close to the star, that applies as long as the energy losses do not dominate the shock-acceleration. 

Figure \ref{fig1} shows that IC over the stellar UV ambient photon field dominates electron energy losses. The X-ray photon spectrum results from folding the electron differential energy spectrum $dN/d\gamma$ with the single electron IC power in the Thomson regime (see Eq.\ref{t_loss_Thom}). At low electron energies ($< 10^5$ keV) and at time scales of tens of minutes (see Fig.\ref{fig1}), the acceleration is faster than IC cooling; thus, the electron energy distribution is steadily replenished by freshly accelerated particles. At time scale of $\sim$ hours, as the number of $\sim$ GeV electrons increases, the IC cooling hampers further acceleration. Therefore, the electron energy spectrum cannot be assumed to be strictly in steady state at any time. However, since at time scale of a few days $t_{acc} \sim t^T_{IC}$, a near-balance between the two processes can be assumed.

{The pulsating shock emission from the star is not expected to affect the spectral shape due to the short cooling time by IC compared with the stellar pulsation time: electrons advected downstream of a shock loose most of their energy to the UV photons in the radiation field before encountering the subsequent shock \citep{Chen.White:91a}. However, ions energy spectrum is not affected by energy losses causing a build-up of high energy ions population in the wind due to crossing of multiple shocks travelling outward. Such ion population might modify the turbulence in the wind at a variety of spatial scales and excite turbulence modes that increase the diffusion for ions accelerated by the subsequent shocks. Ions acceleration and turbulence excitation are not included herein.}

\subsection{IC photon energy spectrum through the stellar corona}

Over the narrow energy range ($\sim$ 1 decade) considered herein, the differential electron spectrum $dN/d\gamma$ can be  approximated by a power law, i.e., $dN/d\gamma = C \gamma^{-\delta}$ with a normalization constant $C$. A power law component is also suggested by the observed energy spectral at photon energies $\gtrsim 1$ keV (see Sect.\ref{sec:observations}). The observed photon spectrum is here calculated by folding $dN/d\gamma$ with the single electron IC power off the photon blackbody distribution in the stellar wind $P^T_{IC}$ \citep{Rybicki.Lightman:86} defined in Eq. \ref{P_IC_Thom}.

The UV radiation field scattered photons and the energetic electrons are assumed to be isotropically distributed; if isotropy of electrons is not altered to the lowest order by the slow wind motion, the radially expanding wind UV field has to be isotropic \citep{Chen.White:91a}. In the Thomson scattering limit ($\gamma \epsilon_0  \ll m c^2 $), the IC scattered  power per unit of volume and energy can be written as \citep[eq. 7.31][]{Rybicki.Lightman:86}
\begin{equation}
\frac{dE}{dV dt\, d\epsilon_1} = \frac{3 \pi \sigma_T C}{h^3 c^2} \bar F(\delta) (k_B T)^{(\delta+5)/2} E^{-\frac{\delta -1}{2}} ,
\end{equation}
where $\sigma_T$ is the Thomson cross-section, $h$ the Planck constant and $\bar  F(\delta) \simeq 1$. 
Due to the scaling of the wind temperature $T(R) = T(R_\star) (R_\star / R)^{1/2}$, the photon flux per unit of energy emitted during the shock expansion between two radii  $R_i$ and $R_f$ can be recast as
\begin{eqnarray}
F(E; R_f, R_i) &=& \frac{1}{E} \int_{R_i}^{R_f} \frac{dE}{dV dt\, d\epsilon_1} \left(\frac{R}{d} \right)^2 dR = 
\frac{3 \pi \sigma_T C}{h^3 c^2} \bar F(\delta) (k_B T)^{\frac{\delta+5}{2}}  \frac{R_\star^3}{d^2} \frac{4}{7-\delta}
\left(\frac{R_f}{R_i}\right)^{(7-\delta)/4}
E^{-\frac{\delta +1}{2}} \nonumber\\
&\propto & E^{-\frac{\delta +1}{2}} \quad [\frac{{\rm photons}}{{\rm cm}^2 \, {\rm s} \, {\rm keV}}]    
\end{eqnarray}
where $d$ is the Cepheid-Earth distance.

Thus, for an observed $F(E; R_f, R_i)$ flux per unit energy  fitted by the non-thermal power law component with index $\Gamma$ (see Sect.\ref{sec:observations}), the power law index of the electrons energy distribution is $\delta = 2(\Gamma -1) +1$ \citep{Rybicki.Lightman:86}. In table 2 a thermal/non-thermal fit for the high state yields a reasonable $\chi^2=1.27$ by freezing $\Gamma \sim 2.0$, or $\delta=3.0$ that corresponds to a quite steep distribution of accelerated electrons. A value $\Gamma \sim 1.6$, hence $\delta=2.2$, leads to a comparable  $\chi^2=1.32$. The linear Diffusive Shock Acceleration model links the electron spectral index of the phase space distribution function $s$ ($s=\delta+2$) with the assumed energy-independent shock density compression $r=s/(s-3) \sim 2.5$ (and $r=3.5$ for $\delta = 2.2$). For radiative shocks the density compression depends on the particle energy because the thickness of the downstream cooling region is probed down to distinct distances from the shock at distinct particle energy \citep[e.g.,][]{Krolik.Raymond:85}. This effect is neglected herein due to the near-balance of the cooling with the IC time scale shown in Fig. \ref{fig1}.     

\subsection{Radio synchrotron}

The assumed magnetic field $0.01$ G at a distance $3\, R_\star$ from the star implies a very long time scale for the radio-synchrotron cooling that is therefore a very valuable constraint on the local magnetic field. 

Radio continuum emission at $15$ GHz (VLA) from $\delta$~Cep was reported by \cite{Matthews.etal:20}; the periodicity of such a radio emission is uncertain, hence its association with stellar pulsation. This radio dataset seems to be consistent \citep{Matthews.etal:22} with a time-variable radio emission at a
level of greater than or equal to 10$\%$, but the likelihood of a correlation with the stellar pulsation seems to be very low.

\section{Conclusion} \label{sec:conclusion}

We have presented an updated spectral fit of the thermal and non-thermal components of the enhanced X-ray emission observed by XMM-Newton in $\delta$~Cep \citep{Engle.etal:17}. We propose a scenario of efficient local pre-ionization of the neutral stellar wind of $\delta$~Cep by a travelling pulsation-driven shock; shocks travelling through the same regions at a later time ($\sim t_p$) encounter a very hot and highly ionized plasma and efficiently accelerate electrons to multi-GeV. Our analysis of the acceleration and losses time scales, combined with previous hydrodynamic simulations \citep{Moschou.etal:20}, suggests that IC over the UV background stellar field dominates the energy losses and limit the shock-acceleration only as electrons reached the GeV energy range; the lack of periodic radio counterpart associated with stellar pulsation  supports this scenario. 
The non-thermal emission  powering the high-flux state does not seem to be completely shut off during the low-flux state. 
Although a two-temperature spectral fit is not statistically ruled out, we find that a significant non-thermal component is consistent with inferred parameters of star and circumstellar medium. If confirmed this model lays the ground for the association of Cepheid pulsations with shock-accelerated GeV electrons.   

{Overall,  multiwavelength (at least in X-rays, radio and IR) coverage will help  improve the understanding of the role of stellar pulsations in shaping the circumstellar material. If further X-ray observational campaigns for $\delta$~Cep, or other Cepheids, will identify a periodicity in X-ray flashes that are lagged by a fraction of the pulsation period from an FUV peak and originate in the near-star environment, additional constraints on the shock properties might be within reach. Furthermore, the exquisite spectral and imaging resolution of JWST from $0.7$ to $10$ $\mu$m might enable characterization with unprecedented precision parameters such as size, distance from the star, gas density and chemical composition of the interstellar clouds surrounding galactic Cepheids, leading the way to a better understanding of the particle acceleration in Cepheid environments.} 

\begin{acknowledgments}
{We thank the referee for useful and constructive comments; we also thank Dr. S. P. Moschou for seminal work on this idea and for useful conversations.} FF was supported, in part, by  NASA through Chandra Theory Award Number $TM0-21001X$, $TM6-17001A$ issued by the Chandra X-ray Observatory Center, which is operated by the Smithsonian Astrophysical Observatory for and on behalf of NASA under contract NAS8-03060, by NASA under Grants 80NSSC18K1213 and by NSF under grant 1850774. JJD and NRE were supported by NASA contract NAS8-03060 to the {\it Chandra X-ray Center} and thanks the Director, Pat Slane, for continuing advice and support. 
\end{acknowledgments}

\appendix

The auxiliary functions for bremmstrahlung losses are defined as
\begin{eqnarray}
A_p(E) = \gamma {\rm ln} {\gamma + p} -p/3 + p^3/\gamma^6(2/9 \gamma^2 - 19/675 \gamma p^2 - 0.06 p^4/\gamma),  \qquad 
A_e(E) =  p/\gamma (\gamma - 1) \Phi(E)
\label{tbrem_aux}
\end{eqnarray}
where $\Phi(E)$ is defined in \cite{Haug:04}\footnote{Figures \ref{fig1} and \ref{fig2} include our correction of a typo in Eq. 7 therein, for the function $\Phi (E)$ in the range $E < 100$ keV}, Eq. 7, as piecewise function for distinct electron energy ranges. 
\newpage

\def \apss{{\it Astrophys.\ Sp.\ Sci.}}
\def \aj{{\it The Astronomical Journal}}
\def \apj{{\it The Astrophysical Journal}}
\def \apjl{{\it The Astrophysical Journal Letters}}
\def \apjs{{\it The Astrophysical Journal Supplement Series}}
\def \araa{{\it Annual Review Astronomy \& Astrophysics}}
\def \prc{{\it Physical\ Review\ C}}
\def \aap{{\it Astronomy \& Astrophysics}}
\def \aaps{{\it Astronomy \& Astrophysics Supplement Series}}
\def \gca{{\it Geochim. Cosmochim.\ Act.}}
\def \grl{{\it Geophysical Research Letters}}
\def \jgr{{\it Journal of Geophysical Research}}
\def \mnras{{\it Monthly Notices of the Royal Astronomical Society}}
\def \nat{{\it Nature}}
\def \physscr{{\it Physica\ Scripta}}
\def \pre{{\it Physical\ Review\ E}}
\def \physrep{{\it Physical\ Report}}
\def \planss{{\it Planetary and Space Science}}
\def \pasp{{\it Publ.\ Astron.\ Soc.\ Pac.}}
\def \pasj{{\it Publications of the Astronomical Society of Japan}}
\def \solphys{{\it Solar\ Physics}}
\def \ssr{{\it Space\ Science\ Reviews}}

\bibliography{ffraschetti}

\begin{thebibliography}{}
\expandafter\ifx\csname natexlab\endcsname\relax\def\natexlab#1{#1}\fi
\providecommand{\url}[1]{\href{#1}{#1}}
\providecommand{\dodoi}[1]{doi:~\href{http://doi.org/#1}{\nolinkurl{#1}}}
\providecommand{\doeprint}[1]{\href{http://ascl.net/#1}{\nolinkurl{http://ascl.net/#1}}}
\providecommand{\doarXiv}[1]{\href{https://arxiv.org/abs/#1}{\nolinkurl{https://arxiv.org/abs/#1}}}

\bibitem[{{Arnaud}(1996)}]{xspec}
{Arnaud}, K.~A. 1996, in Astronomical Society of the Pacific Conference Series,
  Vol. 101, Astronomical Data Analysis Software and Systems V, ed. G.~H.
  {Jacoby} \& J.~{Barnes}, 17

\bibitem[{{Barron} {et~al.}(2022){Barron}, {Wade}, {Evans}, {Folsom}, \&
  {Neilson}}]{Barron.etal:22}
{Barron}, J.~A., {Wade}, G.~A., {Evans}, N.~R., {Folsom}, C.~P., \& {Neilson},
  H.~R. 2022, \mnras, 512, 4021, \dodoi{10.1093/mnras/stac565}

\bibitem[{{Bohm-Vitense} \& {Love}(1994)}]{Bohm-Vitense.Love:94}
{Bohm-Vitense}, E., \& {Love}, S.~G. 1994, \apj, 420, 401,
  \dodoi{10.1086/173570}

\bibitem[{{Chen} \& {White}(1991{\natexlab{a}})}]{Chen.White:91a}
{Chen}, W., \& {White}, R.~L. 1991{\natexlab{a}}, \apj, 366, 512,
  \dodoi{10.1086/169586}

\bibitem[{{Chen} \& {White}(1991{\natexlab{b}})}]{Chen.White:91b}
---. 1991{\natexlab{b}}, \apjl, 381, L63, \dodoi{10.1086/186197}

\bibitem[{{Cleeves} {et~al.}(2017){Cleeves}, {Bergin}, {{\"O}berg}, {Andrews},
  {Wilner}, \& {Loomis}}]{Cleeves.etal:17}
{Cleeves}, L.~I., {Bergin}, E.~A., {{\"O}berg}, K.~I., {et~al.} 2017, \apjl,
  843, L3, \dodoi{10.3847/2041-8213/aa76e2}

\bibitem[{{Colgan} {et~al.}(2008){Colgan}, {Abdallah}, {Sherrill}, {Foster},
  {Fontes}, \& {Feldman}}]{Colgan.etal:08}
{Colgan}, J., {Abdallah}, J., J., {Sherrill}, M.~E., {et~al.} 2008, \apj, 689,
  585, \dodoi{10.1086/592561}

\bibitem[{{David} {et~al.}(2022){David}, {Fraschetti}, {Giacalone},
  {Wimmer-Schweingruber}, {Berger}, \& {Lario}}]{David.etal:22}
{David}, L., {Fraschetti}, F., {Giacalone}, J., {et~al.} 2022, \apj, 928, 66,
  \dodoi{10.3847/1538-4357/ac54af}

\bibitem[{{Degenaar} {et~al.}(2018){Degenaar}, {Ballantyne}, {Belloni},
  {Chakraborty}, {Chen}, {Ji}, {Kretschmar}, {Kuulkers}, {Li}, {Maccarone},
  {Malzac}, {Zhang}, \& {Zhang}}]{Degenaar.etal:18}
{Degenaar}, N., {Ballantyne}, D.~R., {Belloni}, T., {et~al.} 2018, \ssr, 214,
  15, \dodoi{10.1007/s11214-017-0448-3}

\bibitem[{{Drake} {et~al.}(2016){Drake}, {Delgado}, {Laming}, {Starrfield},
  {Kashyap}, {Orlando}, {Page}, {Hernanz}, {Ness}, {Gehrz}, {van Rossum}, \&
  {Woodward}}]{Drake.etal:16b}
{Drake}, J.~J., {Delgado}, L., {Laming}, J.~M., {et~al.} 2016, \apj, 825, 95,
  \dodoi{10.3847/0004-637X/825/2/95}

\bibitem[{{Drake} \& {Linsky}(1986)}]{Drake.Linsky:86}
{Drake}, S.~A., \& {Linsky}, J.~L. 1986, \aj, 91, 602, \dodoi{10.1086/114043}

\bibitem[{{Drury}(1983)}]{Drury:83}
{Drury}, L.~O. 1983, Reports on Progress in Physics, 46, 973,
  \dodoi{10.1088/0034-4885/46/8/002}

\bibitem[{{Drury}(2011)}]{Drury:11}
---. 2011, \mnras, 415, 1807, \dodoi{10.1111/j.1365-2966.2011.18824.x}

\bibitem[{{Engle} {et~al.}(2017){Engle}, {Guinan}, {Harper}, {Cuntz}, {Remage
  Evans}, {Neilson}, \& {Fawzy}}]{Engle.etal:17}
{Engle}, S.~G., {Guinan}, E.~F., {Harper}, G.~M., {et~al.} 2017, \apj, 838, 67,
  \dodoi{10.3847/1538-4357/aa6159}

\bibitem[{{Engle} {et~al.}(2014){Engle}, {Guinan}, {Harper}, {Neilson}, \&
  {Remage Evans}}]{Engle.etal:14}
{Engle}, S.~G., {Guinan}, E.~F., {Harper}, G.~M., {Neilson}, H.~R., \& {Remage
  Evans}, N. 2014, \apj, 794, 80, \dodoi{10.1088/0004-637X/794/1/80}

\bibitem[{{Evans} {et~al.}(2018){Evans}, {Engle}, {Guinan}, {Neilson},
  {Marengo}, {Matthews}, \& {Guenther}}]{Evans.etal:18}
{Evans}, N.~R., {Engle}, S., {Guinan}, E., {et~al.} 2018, in The RR Lyrae 2017
  Conference. Revival of the Classical Pulsators: from Galactic Structure to
  Stellar Interior Diagnostics, ed. R.~{Smolec}, K.~{Kinemuchi}, \& R.~I.
  {Anderson}, Vol.~6, 253--257

\bibitem[{{Evans} {et~al.}(2021){Evans}, {Pillitteri}, {Kervella}, {Engle},
  {Guinan}, {G{\"u}nther}, {Wolk}, {Neilson}, {Marengo}, {Matthews}, {Moschou},
  {Drake}, {Guzik}, {Gallenne}, {M{\'e}rand}, \& {Hocd{\'e}}}]{Evans.etal:21}
{Evans}, N.~R., {Pillitteri}, I., {Kervella}, P., {et~al.} 2021, \aj, 162, 92,
  \dodoi{10.3847/1538-3881/ac05cd}

\bibitem[{{Fraschetti}(2013)}]{Fraschetti:13}
{Fraschetti}, F. 2013, \apj, 770, 84, \dodoi{10.1088/0004-637X/770/2/84}

\bibitem[{{Fraschetti}(2021)}]{Fraschetti:21}
---. 2021, \apj, 909, 42, \dodoi{10.3847/1538-4357/abd699}

\bibitem[{{Fraschetti} \& {Giacalone}(2012)}]{Fraschetti.Giacalone:12}
{Fraschetti}, F., \& {Giacalone}, J. 2012, \apj, 755, 114,
  \dodoi{10.1088/0004-637X/755/2/114}

\bibitem[{{Fraschetti} {et~al.}(2018){Fraschetti}, {Katsuda}, {Sato},
  {Jokipii}, \& {Giacalone}}]{Fraschetti.etal:18}
{Fraschetti}, F., {Katsuda}, S., {Sato}, T., {Jokipii}, J.~R., \& {Giacalone},
  J. 2018, Physical Review Letters, 120, 251101,
  \dodoi{10.1103/PhysRevLett.120.251101}

\bibitem[{{Fraschetti} \& {Pohl}(2017)}]{Fraschetti.Pohl:17b}
{Fraschetti}, F., \& {Pohl}, M. 2017, in European Physical Journal Web of
  Conferences, Vol. 136, 02009, \dodoi{10.1051/epjconf/201713602009}

\bibitem[{{Gallenne} {et~al.}(2021){Gallenne}, {M{\'e}rand}, {Kervella},
  {Pietrzy{\'n}ski}, {Gieren}, {Hocd{\'e}}, {Breuval}, {Nardetto}, \&
  {Lagadec}}]{Gallenne.etal:21}
{Gallenne}, A., {M{\'e}rand}, A., {Kervella}, P., {et~al.} 2021, \aap, 651,
  A113, \dodoi{10.1051/0004-6361/202140350}

\bibitem[{{Giacalone} \& {Jokipii}(2007)}]{Giacalone.Jokipii:07}
{Giacalone}, J., \& {Jokipii}, J.~R. 2007, \apjl, 663, L41,
  \dodoi{10.1086/519994}

\bibitem[{{Gillet}(2014)}]{Gillet:14}
{Gillet}, D. 2014, \aap, 568, A72, \dodoi{10.1051/0004-6361/201423486}

\bibitem[{{Hampel} {et~al.}(2022){Hampel}, {Komossa}, {Greiner}, {Reiprich},
  {Freyberg}, \& {Erben}}]{Hampel.etal:22}
{Hampel}, J., {Komossa}, S., {Greiner}, J., {et~al.} 2022, Research in
  Astronomy and Astrophysics, 22, 055004, \dodoi{10.1088/1674-4527/ac5800}

\bibitem[{{Haug}(2004)}]{Haug:04}
{Haug}, E. 2004, \aap, 423, 793, \dodoi{10.1051/0004-6361:20040377}

\bibitem[{{Hocd{\'e}} {et~al.}(2020){Hocd{\'e}}, {Nardetto}, {Lagadec},
  {Niccolini}, {Domiciano de Souza}, {M{\'e}rand}, {Kervella}, {Gallenne},
  {Marengo}, {Trahin}, {Gieren}, {Pietrzy{\'n}ski}, {Borgniet}, {Breuval}, \&
  {Javanmardi}}]{Hocde.etal:20}
{Hocd{\'e}}, V., {Nardetto}, N., {Lagadec}, E., {et~al.} 2020, \aap, 633, A47,
  \dodoi{10.1051/0004-6361/201935848}

\bibitem[{{Inoue} {et~al.}(2012){Inoue}, {Yamazaki}, {Inutsuka}, \&
  {Fukui}}]{Inoue.etal:12}
{Inoue}, T., {Yamazaki}, R., {Inutsuka}, S.-i., \& {Fukui}, Y. 2012, \apj, 744,
  71, \dodoi{10.1088/0004-637X/744/1/71}

\bibitem[{{Jokipii}(1982)}]{Jokipii:82}
{Jokipii}, J.~R. 1982, \apj, 255, 716, \dodoi{10.1086/159870}

\bibitem[{{Jokipii}(1987)}]{Jokipii:87}
---. 1987, \apj, 313, 842, \dodoi{10.1086/165022}

\bibitem[{{Kervella} {et~al.}(2006){Kervella}, {M{\'e}rand}, {Perrin}, \&
  {Coud{\'e} du Foresto}}]{Kervella.etal:06}
{Kervella}, P., {M{\'e}rand}, A., {Perrin}, G., \& {Coud{\'e} du Foresto}, V.
  2006, \aap, 448, 623, \dodoi{10.1051/0004-6361:20053603}

\bibitem[{{Krolik} \& {Raymond}(1985)}]{Krolik.Raymond:85}
{Krolik}, J.~H., \& {Raymond}, J.~C. 1985, \apj, 298, 660,
  \dodoi{10.1086/163650}

\bibitem[{{Kuntz} \& {Snowden}(2008)}]{kuntz08}
{Kuntz}, K.~D., \& {Snowden}, S.~L. 2008, \aap, 478, 575,
  \dodoi{10.1051/0004-6361:20077912}

\bibitem[{{Landi} \& {Landini}(1999)}]{Landi.Landini:99}
{Landi}, E., \& {Landini}, M. 1999, \aap, 347, 401

\bibitem[{{Longair}(1994)}]{Longair:94}
{Longair}, M.~S. 1994, {High energy astrophysics. Volume 2. Stars, the Galaxy
  and the interstellar medium.}, Vol.~2

\bibitem[{{Lucy}(1982{\natexlab{a}})}]{Lucy:82a}
{Lucy}, L.~B. 1982{\natexlab{a}}, \apj, 255, 278, \dodoi{10.1086/159826}

\bibitem[{{Lucy}(1982{\natexlab{b}})}]{Lucy:82b}
---. 1982{\natexlab{b}}, \apj, 255, 286, \dodoi{10.1086/159827}

\bibitem[{{Lucy} \& {Solomon}(1970)}]{Lucy.Solomon:70}
{Lucy}, L.~B., \& {Solomon}, P.~M. 1970, \apj, 159, 879, \dodoi{10.1086/150365}

\bibitem[{{Lucy} \& {White}(1980)}]{Lucy.White:80}
{Lucy}, L.~B., \& {White}, R.~L. 1980, \apj, 241, 300, \dodoi{10.1086/158342}

\bibitem[{{Markowitz} {et~al.}(2022){Markowitz}, {Nalewajko}, {Bhatta},
  {Dewangan}, {Chandra}, {Dorner}, {Schleicher}, {Pajdosz-{\'S}mierciak},
  {Stawarz}, {Zola}, {Ostrowski}, {Carosati}, {Krishnan}, {Bachev},
  {Ben{\'\i}tez}, {Gazeas}, {Hiriart}, {Hu}, {Larionov}, {Marchini},
  {Matsumoto}, {Nikiforova}, {Pursimo}, {Raiteri}, {Reichart}, {Rodriguez},
  {Semkov}, {Strigachev}, {Sugiura}, {Villata}, {Webb}, {Arbet-Engels},
  {Baack}, {Balbo}, {Biland}, {Bretz}, {Buss}, {Eisenberger}, {Elsaesser},
  {Hildebrand}, {Iotov}, {Kalenski}, {Mannheim}, {Mitchell}, {Neise}, {Noethe},
  {Paravac}, {Rhode}, {Sliusar}, \& {Walter}}]{Markowitz.etal:22}
{Markowitz}, A.~G., {Nalewajko}, K., {Bhatta}, G., {et~al.} 2022, \mnras, 513,
  1662, \dodoi{10.1093/mnras/stac917}

\bibitem[{{Mathias} {et~al.}(2006){Mathias}, {Gillet}, {Fokin}, {Nardetto},
  {Kervella}, \& {Mourard}}]{Mathias.etal:06}
{Mathias}, P., {Gillet}, D., {Fokin}, A.~B., {et~al.} 2006, \aap, 457, 575,
  \dodoi{10.1051/0004-6361:20065299}

\bibitem[{{Matthews} {et~al.}(2022){Matthews}, {Evans}, \&
  P.}]{Matthews.etal:22}
{Matthews}, L.~D., {Evans}, N.~R., \& P., R.~M. 2022, \aj, submitted

\bibitem[{{Matthews} {et~al.}(2020){Matthews}, {Evans}, \&
  {Rupen}}]{Matthews.etal:20}
{Matthews}, L.~D., {Evans}, N.~R., \& {Rupen}, M.~P. 2020, in American
  Astronomical Society Meeting Abstracts, Vol. 235, American Astronomical
  Society Meeting Abstracts \#235, 106.03

\bibitem[{{Matthews} {et~al.}(2012){Matthews}, {Marengo}, {Evans}, \&
  {Bono}}]{Matthews.etal:12}
{Matthews}, L.~D., {Marengo}, M., {Evans}, N.~R., \& {Bono}, G. 2012, \apj,
  744, 53, \dodoi{10.1088/0004-637X/744/1/53}

\bibitem[{{Meinecke} {et~al.}(2014){Meinecke}, {Doyle}, {Miniati}, {Bell},
  {Bingham}, {Crowston}, {Drake}, {Fatenejad}, {Koenig}, {Kuramitsu},
  {C.~Kuranz}, {Lamb}, {Lee}, {MacDonald}, {Murphy}, {Park}, {Pelka},
  {Ravasio}, {Sakawa}, {Schekochihin}, {Scopatz}, {Tzeferacos}, {Wan},
  {Woolsey}, {Yurchak}, {Reville}, \& {Gregori}}]{Meinecke.etal:14}
{Meinecke}, J., {Doyle}, H.~W., {Miniati}, F., {et~al.} 2014, Nature Physics,
  10, 520, \dodoi{10.1038/nphys2978}

\bibitem[{{M{\'e}rand} {et~al.}(2006){M{\'e}rand}, {Kervella}, {Coud{\'e} du
  Foresto}, {Perrin}, {Ridgway}, {Aufdenberg}, {ten Brummelaar}, {McAlister},
  {Sturmann}, {Sturmann}, {Turner}, \& {Berger}}]{Merand.etal:06}
{M{\'e}rand}, A., {Kervella}, P., {Coud{\'e} du Foresto}, V., {et~al.} 2006,
  \aap, 453, 155, \dodoi{10.1051/0004-6361:20054466}

\bibitem[{{Moschou} {et~al.}(2020){Moschou}, {Vlahakis}, {Drake}, {Evans},
  {Neilson}, {Guzik}, \& {ZuHone}}]{Moschou.etal:20}
{Moschou}, S.-P., {Vlahakis}, N., {Drake}, J.~J., {et~al.} 2020, \apj, 900,
  157, \dodoi{10.3847/1538-4357/aba8fa}

\bibitem[{{Nardetto} {et~al.}(2006){Nardetto}, {Mourard}, {Kervella},
  {Mathias}, {M{\'e}rand}, \& {Bersier}}]{Nardetto.etal:06}
{Nardetto}, N., {Mourard}, D., {Kervella}, P., {et~al.} 2006, \aap, 453, 309,
  \dodoi{10.1051/0004-6361:20054333}

\bibitem[{{Owocki}(2015)}]{Owocki:15}
{Owocki}, S.~P. 2015, in Astrophysics and Space Science Library, Vol. 412, Very
  Massive Stars in the Local Universe, ed. J.~S. {Vink}, 113,
  \dodoi{10.1007/978-3-319-09596-7_5}

\bibitem[{{Parizot} {et~al.}(2006){Parizot}, {Marcowith}, {Ballet}, \&
  {Gallant}}]{Parizot.etal:06}
{Parizot}, E., {Marcowith}, A., {Ballet}, J., \& {Gallant}, Y.~A. 2006, \aap,
  453, 387, \dodoi{10.1051/0004-6361:20064985}

\bibitem[{{Rybicki} \& {Lightman}(1986)}]{Rybicki.Lightman:86}
{Rybicki}, G.~B., \& {Lightman}, A.~P. 1986, {Radiative Processes in
  Astrophysics}

\bibitem[{{Sasselov} \& {Lester}(1994)}]{Sasselov.Lester:94}
{Sasselov}, D.~D., \& {Lester}, J.~B. 1994, \apj, 423, 795,
  \dodoi{10.1086/173858}

\bibitem[{{Sato} {et~al.}(2018){Sato}, {Katsuda}, {Morii}, {Bamba}, {Hughes},
  {Maeda}, {Ishida}, \& {Fraschetti}}]{Sato.etal:18}
{Sato}, T., {Katsuda}, S., {Morii}, M., {et~al.} 2018, \apj, 853, 46,
  \dodoi{10.3847/1538-4357/aaa021}

\bibitem[{{Schlickeiser}(2009)}]{SChlickeiser:09}
{Schlickeiser}, R. 2009, \mnras, 398, 1483,
  \dodoi{10.1111/j.1365-2966.2009.15205.x}

\bibitem[{{Scowcroft} {et~al.}(2016){Scowcroft}, {Seibert}, {Freedman},
  {Beaton}, {Madore}, {Monson}, {Rich}, \& {Rigby}}]{Scowcroft.etal:16}
{Scowcroft}, V., {Seibert}, M., {Freedman}, W.~L., {et~al.} 2016, \mnras, 459,
  1170, \dodoi{10.1093/mnras/stw628}

\bibitem[{{Uchiyama} \& {Aharonian}(2008)}]{Uchiyama.Aharonian:08}
{Uchiyama}, Y., \& {Aharonian}, F.~A. 2008, \apjl, 677, L105,
  \dodoi{10.1086/588190}

\bibitem[{{Wilkinson} \& {Uttley}(2009)}]{Wilkinson.Uttley:09}
{Wilkinson}, T., \& {Uttley}, P. 2009, \mnras, 397, 666,
  \dodoi{10.1111/j.1365-2966.2009.15008.x}

\end{thebibliography}

\end{document}